\documentclass[aps,prb,reprint,superscriptaddress]{revtex4-2}
\usepackage{graphicx}
\usepackage{amsmath}
\usepackage{bm}  
\usepackage{chemformula} 
\usepackage{multirow}
\usepackage{geometry}
\usepackage{lastpage}
\usepackage[utf8]{inputenc}
\usepackage[T1]{fontenc}
\usepackage[justification=Justified]{caption}
\usepackage{float}
\usepackage[english]{babel}
\addto{\captionsenglish}{}
\usepackage{url} 
\usepackage{hyperref}
\usepackage{epstopdf}

\begin{document}
\title{Multi-Functional Properties of Manganese Pnictides: A First-Principles Study on Magneto-Optics and Magnetocaloric Properties}

\author{S Jayendran}
\affiliation{Department of Physics, Central University of Tamil Nadu, Thiruvarur 610005, Tamil Nadu, India}
\affiliation{Simulation Center for Atomic and Nanoscale MATerials(SCANMAT), Central University of Tamil Nadu, Thiruvarur 610005, Tamil Nadu, India}
\author{K G Abhishek}
\affiliation{Department of Physics, Central University of Tamil Nadu, Thiruvarur 610005, Tamil Nadu, India}
\affiliation{Simulation Center for Atomic and Nanoscale MATerials(SCANMAT), Central University of Tamil Nadu, Thiruvarur 610005, Tamil Nadu, India}
\author{R Suresh}
\affiliation{Simulation Center for Atomic and Nanoscale MATerials(SCANMAT), Central University of Tamil Nadu, Thiruvarur 610005, Tamil Nadu, India}
\author{Helmer Fjellv\aa g}
\affiliation{Department of Chemistry, University of Oslo, P.O. Box 1033 Blindern, N-0315 Oslo, Norway}
\author{P Ravindran}
\email{raviphy@cutn.ac.in}
\affiliation{Department of Physics, Central University of Tamil Nadu, Thiruvarur 610005, Tamil Nadu, India}
\affiliation{Simulation Center for Atomic and Nanoscale MATerials(SCANMAT), Central University of Tamil Nadu, Thiruvarur 610005, Tamil Nadu, India}

\begin{abstract}
Magnetic refrigeration presents an energy-efficient and environmentally benign alternative to traditional vapour-compression cooling technologies. It relies on the magnetocaloric effect, in which the temperature of a magnetic material changes in response to variations in an applied magnetic field. Optimal magnetocaloric materials are characterized by a significant change in magnetic entropy under moderate magnetic field. In this study, we systematically investigated the inter-atomic exchange interactions, magnetic anisotropy energy and magnetocaloric properties of Mn$X$ ($X$ = N, P, As, Sb, Bi) using a combination of density functional theory and Monte-Carlo simulations. Additionally, the magneto-optical Kerr and Faraday spectra were computed using the all-electron, fully relativistic, full-potential linearized muffin-tin orbital method. The largest Kerr effect observed in MnBi can be inferred as a combined effect of maximal exchange splitting of Mn 3$d$ states and the large spin-orbit coupling of Bi. To extract site-projected spin and orbital moments, spin-orbit coupling and orbital polarization correction are accounted in the present calculation, which shows good agreement between the moment obtained from the X-ray magnetic circular dichroism sum rule analysis, spin-polarized calculation, and experimental studies. The magnetic transition temperatures predicted through Monte-Carlo simulations were in good agreement with the corresponding experimental values. Our results provide a unified microscopic understanding of magnetocaloric performance and magneto-optical activity in Mn-based pnictides and establish a reliable computational framework for designing next-generation magnetic refrigeration materials.	
\end{abstract}

\maketitle
\section{Introduction}
Recent issues concerning climate changes encourage the development of technologies which are more energy efficient and environmentally friendly. One of the major sources of energy consumption is refrigeration and air conditioning. Conventional gas compression-expansion refrigeration techniques are highly energy consuming and causing global warming by emitting greenhouse gases (GHG) such as chlorofluorocarbon (CFC) and hydrofluorocarbon (HFC) which causes ozone layer depletion, further accelerating the rise in global temperature. The poor efficiency of traditional air conditioners further give strain to the environment. In this scenario more studies are directed towards efficient energy utilization and reduction in GHG emissions. Solid state refrigerators are a major break through in this area and magnetocaloric materials are one among them. These materials work on the phenomena known as the magnetocaloric effect (MCE). The MCE was discovered in 1881 by Emil Warburg~\cite{warburg}. It is the temperature change induced due to a magnetic material's response to an externally applied magnetic field. For a simple ferromagnet near its Curie temperature (T${_C}$), when a magnetic field is applied, the randomly oriented spins are aligned parallel to the applied magnetic field. This transforms the system from a less ordered state to a more ordered state that lowers magnetic entropy $\Delta S_M$. Hence, to compensate the total entropy the lattice entropy of the system increases. In other words, the molecules of the system start to vibrate more frequently and the temperature of the system rises. When external magnetic field is withdrawn, the system reverts to the initial state and the temperature falls down. The MCE is used for several applications, among which magnetic refrigeration is the prominent one. 

Debye (1926)\cite{debye} and Giauque (1927)\cite{giauque} independently devised the concept of cooling through magnetic field fluctuation, calling it adiabatic demagnetization. This domain started to emerge after the development of a magnetic refrigerator using metallic gadolinium as a magnetic refrigerant by Brown\citep{brown1976} in 1976. The major breakthrough happened after 20 years, in 1997  when Pecharsky \& Gschneidner\cite{gschneidner} discovered giant magnetocaloric effect in Gd$_{5}$Si$_{2}$Ge$_{2}$ and related compounds. They have shown that the peak in the isothermal entropy change of 20 J/(kg K) in $Gd_5Si_2Ge_2$ is happening around 273 K upon magnetic field variation from 0 to 5\,T. The magnetocaloric effect is associated to a large change in magnetization in the vicinity of working temperature of the refrigerant material. The low temperature refrigerants rely on paramagnetic salts, due to remarkable increase in the magnetization as it approaches absolute zero. However, a different approach had to be adopted for room temperature application where one will expect a phase transition in the material closer to working (room) temperature.

Magnetic phase transitions are classified into first and second order transitions. First order transition is characterised by an abrupt change in magnetization at the phase transition point (i.e. a discontinuous change in first derivative of Gibbs free energy). The large change in magnetization during the transition can cause a giant magnetic entropy change. In addition to that thermal as well as magnetic hysteresis also appear and these should be avoided for efficient refrigerator application. In second order phase transition, a lack of thermal and magnetic hysteresis results in a continuous decrease of magnetization to zero. The higher value of magnetic moment plays an important role in enhancing MCE and as a consequence of that, until 2007 the magnetocaloric effect has been mostly studied in rare earth metals and their alloys\cite{gschneidner}. One of the major disadvantages of these materials are their cost, availability and low transition temperature. Hence, it cannot be used for commercial purposes. This, in turn, leads to the search for magnetocaloric materials based on magnetic 3$d$ transition-metal elements, which are significantly less expensive than those based on rare earth elements. However, the atomic moments of 3$d$ elements are much smaller than those of the rare earth elements. In intermetallic compounds, the magnetic moments of Ni, Co, and Fe are below 0.6, 1.7 and 2.2 $\mu _B$/atom respectively. In contrast, much higher values can be reached in Mn-based intermetallics, even 4 $\mu _B$/Mn atom. So, it is more attractive to study Mn based compounds for practical MCE applications.\cite{MnP_MCE_2009,mnas_mce,MnSb,mnbi_mce}

\begin{figure}[h]
	\centering
	\includegraphics[height=3cm]{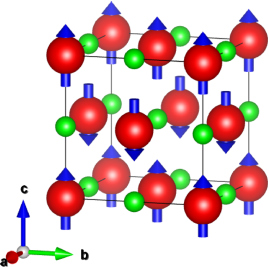}
	\includegraphics[height=3cm]{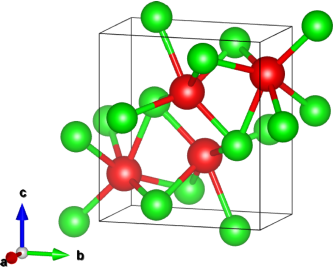}
	\hfill
    \includegraphics[height=3cm]{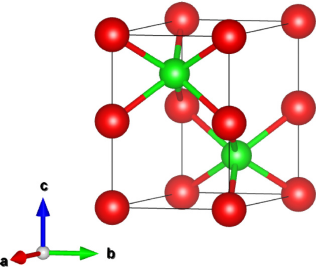}
	\caption{Optimized equilibrium crystal structures of Mn$X$ ($X$ = N, P, As, Sb, Bi) compounds: (right panel) MnN in the A-type antiferromagnetic configuration, (middle panel) orthorhombic MnP, and (left panel) hexagonal MnAs/MnSb/MnBi. Red and green spheres represent Mn and $X$ atoms, respectively.}
	\label{fig:struc}
\end{figure}

Mn$X$ ($X$ = N, P, As, Sb, Bi) type of materials have received considerable research interest because of their remarkable magnetocaloric and magneto optical (MO) properties. Over the past two to three decades, manganese pnictides such as MnP \cite{ManganesePnictidesMnP,HelicalMagneticState,MnP-mae,MnP_MCE_2009,tranTheoreticalPredictionLarge2021}, MnAs\cite{MnAs2000,MnAs2004,MnAs2007,MnAs2010,MnAs2014,ravindranMnX}, MnSb \cite{MnSb1976,MnSb2015,MnSbMagnetocrystalline1992,MnSbinter2015,MnSbstudiesSpinwave1996}, and MnBi \cite{MnBi1955,MnBi2014,MnBi2022,MnBiMagneticStructuralPhase2008,MnBieutronDiffractionStudy1956,MnBijij2016} are extensively studied using both experimental and theoretical approaches. All the materials except MnN in this series are ferromagnetic with Curie temperature of MnP, MnAs, MnSb, and MnBi being 290\,K\cite{MnP_MCE_2009}, 318\,K\cite{MnAs}, 583\,K\cite{MnSbstudiesSpinwave1996} and 628\,K\cite{MnBiMagneticStructuralPhase2008}, respectively. However, the ground state of MnN is antiferromagnetic with N\'eel temperature of 660\,K\cite{MnN_neel} observed experimentally. In addition to remarkable MCE, manganese pnictides are also attractive for their strong MO activity. There has been a lot of experimental and theoretical efforts to understand the origin of MO property of these materials.\cite{buschow,sabiryanov,huang,Kulatov1995,sato_bulk_mnsb,fumagalli1996,kohler1997calculated,harder,ravindranMnX,shu,oppeneer2001magneto}. In this series, MnAs\cite{ikekame} exhibits the Kerr rotation of $-$0.5$^\circ$ at 1.75\,eV which increases to $-$1.6$^\circ$ at 1.8\,eV for MnBi,\cite{Di_1992} primarily due to the progressive enhancement of spin-orbit coupling (SOC) with increasing atomic number of the pnictogen atoms. For lighter compounds such as MnN and MnP there is no experimental/theoretical MO studies available in the literature. So, the systematic studies in the variation of pnictogen will give more insight into the role of SOC to MO properties of these systems. The large spin moment on Mn site and the strong SOC of heavier pnictogens play a decisive role in determining both the MO activity and magnetic anisotropy. X-ray absorption spectroscopy (XAS) and X-ray magnetic circular dichroism (XMCD) have been extensively employed to probe element-specific magnetic moment in magnetic materials. Such studies are already performed experimentally for Mn$X$ thin-film systems\cite{mnn_xmcd,mnp_xmcd,mnas_xmcd,mnsb_xmcd,mnbi_xmcd}, whereas for bulk samples such studies remain scarce. Since the Faraday spectra requires thin films with high optical transparency, the experimental Faraday spectra measurements are very much restricted to these systems. The hybridization between Mn-3$d$ and X-$p$ states strongly influences their electronic structure and exchange interactions, making Mn$X$ compounds ideal candidates for exploring the interplay between bonding with magnetism, SOC and optical transitions. In order to understand the variation in the SOC, bonding interaction of transition metal with pnictogens, magnetic moment at Mn site, exchange interactions, MO responses, MCE and MAE, we have systematically investigated these materials. Fully relativistic calculations are employed with full-potential treatment to accurately capture the combined effects of exchange splitting and SOC, to provide valuable insights into the origin of large MO effects in manganese pnictides.

The rest of the article is organized as follows. The computational details regarding structural optimization, calculations of the exchange coupling, magneto-optics and magnetocaloric properties, are given in Sec. II. The crystal structure details are in Sec.III. In Sec. IV, the results from our calculations are presented and discussed. The most important conclusions from our calculations are given in Sec. V.

\section{Computational Details}
\subsection{Structural Optimization}
The projector augmented wave (PAW)~\cite{PAW} method, as implemented in the Vienna \textit{ab-initio} simulation package (VASP)~\cite{kresse1996} were used to obtain the ground-state structural parameters for Mn$X$ ($X$\,=\,N, P, As, Sb, Bi). We have used Perdew-Burke-Ernzerhof’s (PBE) generalized gradient approximation~\cite{perdewGGA1996,perdewGGA1997} to approximate the exchange-correlation potential. An energy cut-off for the plane-wave basis of 600\,eV was used to expand the plane wave basis set which is found to be sufficient to accurately predict the structural parameters. Monkhorst-Pack special \textbf{k}-point method with a \textbf{k}-point grid of $8\times8\times6$ (Hexagonal phase), $4\times8\times6$ (Orthorhombic phase) and $8\times8\times8$ (body-centred tetragonal) within the irreducible Brillouin zone(IBZ) for geometry optimization were used. During the calculations, the shape of the crystal and the ionic positions were relaxed using stress and force minimization, respectively until they attain an energy convergence and the force convergence criterion of 10$^{-7}$\,eV/cell and  1\,meV/\AA, respectively. Optimized ground state crystal structures of MnX are shown in Fig.\ref{fig:struc}.

\subsection{Calculation of Exchange Interaction}
The optimized crystal structure obtained from computational aspects mentioned in the previous section was used for exchange interaction calculation through full-potential linear muffin-tin orbital (FP-LMTO) code RSPt\cite{rspt_book}. Among the considered compounds the calculated Curie temperature for MnSb and MnBi obtained from GGA calculations are severely underestimated irrespective of methodologies adopted such as mean field approximation or more accurate MC simulation. The exchange-correlation energy functional was chosen in the form of GGA+U parametrization with $U$ (where $U$ is the on-site Coulomb interaction). It is well-known that in transition metal compound the strong Coulomb correlation effect is playing an important role to decide the physical properties of the systems. So, in order to account for the Coulomb correlation effect properly we have used the values of $U$=0.9\,eV and $U$=0.7\,eV for MnSb and MnBi, respectively\cite{mnx_u}, to achieve Curie temperatures in reasonable agreement with experimental observations.  Magnetic exchange interactions were estimated using the method of Liechtenstein \textit{et al.}\cite{lichtenstein1987}. The exchange parameters are found out from the total energy variation by the perturbation caused due to small rotation of the interacting electrons and the corresponding change in the electron spin density. For this, the DFT Hamiltonian is mapped onto an effective Heisenberg Hamiltonian with classical spins in the following form. 

\begin{equation}
	H =\sum_{i\neq j}J_{ij} \mathbf{e_i.e_j}
\end{equation}
Here $(i,j)$ are the indices for the magnetic sites and $\textbf{e}_i.\textbf{e}_j$ are the unit vectors along the spin direction at sites $i$ and $j$, respectively. The expression for pair exchange parameter $J_{ij}$ is as follows.

\begin{equation}
	J_{ij} = \frac{T}{4} \sum_{n} 
	\mathrm{Tr}\!\left[
	\hat{\Delta}_{i}(i\omega_{n})
	\hat{G}^{\uparrow}_{ij}(i\omega_{n})
	\hat{\Delta}_{j}(i\omega_{n})
	\hat{G}^{\downarrow}_{ji}(i\omega_{n})
	\right],
\end{equation}
where \textit{T} is the temperature, $\omega_n=2\pi T(2n+1)$ is the $n^{th}$ fermionic Matsubara frequency, $\hat{\Delta}_{i/j}$ is the exchange splitting at site $i/j$, $\hat{G}^{\sigma}_{ij}$ is the intersite Green's function between sites $i$ and $j$ projected over a given spin $\sigma$. A minimum of 5000 \textbf{k}-points was ensured for sampling the full Brillouin zone in exchange interaction calculation. A positive or negative $J_{ij}$ represent ferromagnetic or antiferromagnetic interaction, respectively. The exchange interaction results confirm that all these compounds have a ferromagnetic order except MnN, which has an antiferromagnetic ground state. Based on the calculated $J_{ij}$ the magnetic structure of MnN is resolved and is used for other ground state calculations.

\subsection{Calculation of Magnetic Entropy Change}
The isothermal magnetic entropy change in magnetocaloric materials can be calculated from their magnetization as a function of temperature with applied magnetic field by following thermodynamical Maxwell's relation.

\begin{equation}
	\Delta S_m = \mu_{0} \int_{H_i}^{H_f} \left (\frac{\partial M(T, H)}{\partial T}\right) _H dH \,
\end{equation}

where $H_{i}$ and $H_{f}$ are initial and final magnetic fields, respectively, $\mu_{0}$ is the permeability of free space, $M$ is the magnetization, and $T$ is the temperature. The atomistic spin dynamics package UppASD\cite{UppASD2008} was utilized to obtain the field and temperature dependent magnetization by simulating with magnetic exchange interaction parameters ($J_{ij}$) calculated from \textit{ab-initio} method described above. This involves a super cells with dimension of $20\times20\times20$ running for 5,00,000 steps at each temperature, enabling the in-plane periodic boundary conditions to reach equilibrium. Next, the mean magnetization was extracted by a statistical average obtained over the fore-mentioned steps. Temperature-dependent magnetization $(M vs T)$ curves for different applied magnetic field were generated by systematically increasing the external magnetic field upto 5\,T with a step size of 1\,T.

\subsection{Calculation of optical properties}
To investigate the linear optical properties, we adopted FP-LMTO and APW+lo\cite{wien2k} methods to calculate optical dielectric tensor and unscreened plasma frequency (to account intraband contribution), respectively. For metals, intraband contribution needs to be added to the components of the optical conductivity tensors which influence optical property upto 2\,eV. The momentum  transfer from the initial state to the final state was neglected by adopting dipole approximation in our interband optical transition calculations(i.e. no significant change in electron’s momentum during a transition because the photon’s momentum is so small). The SOC is included in these optical calculations to account spin-flip transition and to break spin degeneracy. The interband contribution to absorptive part of the optical conductivity ($\sigma_{\alpha\beta}^{abs}(\omega)$) as a function of photon frequency($\omega$) within the random phase approximation, without considering the local field effects was calculated by summing all the allowed transitions from occupied states to unoccupied states over the BZ weighted with appropriate matrix element and the corresponding probability of transition can be obtained by 
\begin{align}
	\sigma_{\alpha\beta}^{(\text{abs})}(\omega) 
	&= \frac{V e^2}{8 \pi^2 \hbar m^2 \omega} 
	\sum_{nn'} \int d^3 k\, \langle \mathbf{k}n \ | p_\alpha | \mathbf{k}n' \rangle \nonumber \\
	&\quad \times \langle \mathbf{k}n' | p_\beta | \mathbf{k}n \rangle 
	f_{\mathbf{k}n}(1 - f_{\mathbf{k}n'}) 
	\delta(\epsilon_{\mathbf{k}n'} - \epsilon_{\mathbf{k}n} - \hbar \omega).
	\label{optic}
\end{align}
The intraband contribution to the diagonal components of the optical conductivity is normally described by Drude formula\cite{drude1900},
\begin{align}
	\sigma_{D}(\omega) = \frac{\omega_P^2}{4\pi[(1/\tau)-i\omega]}
\end{align}
where, $\tau$ is the relaxation time for characterizing the scattering of charge carriers, and is dependent on quality of the sample. Here we have used the value of $\hbar/\tau  = 0.2 eV$(As mentioned and used in similar compounds\cite{ravindranMnX} in previous studies). We have calculated the Drude parameters for Mn$X$ by integrating over the Fermi surface using the relation 
\begin{align}
	\sigma_{Pii}^{2}(\omega) 
	&= \frac{8 \pi e^2}{V} \sum_{kn} \langle \mathbf{k}n\xi | p_i | \mathbf{k}n\xi \rangle \langle \mathbf{k}n\xi | p_i | \mathbf{k}n\xi \rangle 	\delta(E_{\mathbf{k}n\xi} - E_{F}).
\end{align}
Where, $V$ is the volume of primitive cell, $E_{F}$ is the Fermi energy, $e$ is the electron charge and $\xi$ is the electron spin. By incorporating the orbital polarization (OP) correction as proposed by Eriksson \textit{et al.}\cite{OP}, Hund’s second rule is effectively restored in the Hamiltonian, enabling a more accurate description of the orbital magnetic moment. For \textbf{k}-points sampling, a linear tetrahedron method with a minimum of 6000 \textbf{k}-points were used for sampling the IBZ. To account for broadening effects in the calculated optical spectra, the absorptive optical conductivity is broadened with a Lorentzian function\cite{SnI2}. The full width at half maximum(FWHM) of the Lorentzian is considered to be increase linearly with energy, being 0.02\,eV at photon energy 1\,eV. The real(dispersive) components of the optical conductivity were calculated with a Kramer-Kronig transformation. Since the Kramers-Kronig relation involves an integration of $\epsilon_{2}(\omega)$ from $0$ to $\infty$, a higher energy cut-off is required. In this work, $\epsilon_{2}(\omega)$ was calculated up to 50 eV, which we found sufficient to ensure convergence of optical and magneto-optical spectra from our previous study\cite{SnI2}. Together with the imaginary parts of the optical conductivity, this allows to calculate the key optical constants.

\subsection{Calculation of Magneto Optical Properties}
The magneto optical Kerr effect (MOKE) and Faraday effect (MOFE) describe the changes in the polarization state of polarized light pass on the magnetized system. Specifically MOKE refers to change in polarization during reflection while MOFE refers to changes occurring upon transmission through the magnetic materials. This includes a rotation of the polarization plane and the emergence of circularly polarized to elliptically polarized light, indicative of the complex dielectric response of the material. These changes are typically accompanied by variations in the reflected light intensity, and together, they serve as sensitive probes of the materials electronic structure and magnetic ordering. The origin of MOKE can be traced to the off-diagonal components of the Eq.\ref{optic} induced by SOC. These off-diagonal components impart an anisotropic permittivity to the material leading to fluctuations in the phase of the incident polarized light upon reflection. MOKE is quantitatively assessed by the Kerr angle, which measures the rotation of circularly polarized light by the material's magnetic field and the Kerr ellipticity, representing the ratio of the semi-major and semi-minor axes of the elliptically polarized light after reflection. For polar geometry, the complex polar Kerr angle is given (in the approximation of small angles) by\cite{kerr}
\begin{equation}
	\Phi_K = \theta_K + i \epsilon_K 
	= \frac{-\sigma_{xy}}{\sigma_{xx} \sqrt{1 + i \left( \frac{4\pi \sigma_{xx}}{\omega} \right)}}
	= \frac{-\sigma_{xy}}{D(\omega)},
\end{equation}
where $\theta_K$ and $\epsilon_K$ are the Kerr rotation angle and ellipticity, respectively and $\omega$ is the incident photon energy, $D(\omega)$ is the denominator of the Kerr angle, containing diagonal terms of optical conductivity. The Faraday effect is a phenomenon wherein the plane of polarization of circularly polarized light rotates as it propagates through a magnetized material. This rotation arises from circular birefringence, whereby left and right-circularly polarized light components experience different refractive indices, resulting in a phase difference that manifests as a rotation of the linear polarization plane. This effect is linearly proportional to both the strength of the magnetic field and the optical path length within the material. It may be noted that MOFE is observed only when the magnetic field is aligned parallel to the direction of propagation of light. Which is quantitatively assessed by the Faraday angle, that measures the rotation of circularly polarized light and the Faraday ellipticity. In magnetic circular dichroism, Faraday ellipticity ($\epsilon_{F}$) is proportional to the difference in absorption of right and left circularly polarized light, which is given by\cite{fara}
\begin{equation}
	\theta_F + i \epsilon_F 
	= \frac{\omega d}{2c}(n_+ - n_-),
	\label{fara_equ}
\end{equation}
where $c$ is the velocity of the light in vacuum and $d$ is the thickness of the magnetic thin film.

\section{Crystal Structure Details}
The manganese pnictides mainly crystallizes in $B8_1$ hexagonal(\textit{P6$_3$/mmc}, No. 194) ($X$ = As, Sb, Bi) and $B3_1$ orthorhombic(\textit{Pnma}, No. 62) ($X $= P) structures in low temperature otherwise called as MnP-type and NiAs-type structures, respectively. Among them MnN compound stabilizes in NaCl-type structure(\textit{Fm-3m}, No. 225) with slight tetragonal distortion (so called $\theta$ phase with $c/a=0.98$). The NiAs-type structure has a hexagonal symmetry that possess 4 atoms in the unit cell those are in high symmetric positions, whereas, the orthorhombic $B3_1$ structure is having lower symmetry with relatively larger unit cell that having 8 atoms per cell. In contrast, the MnN system stabilizes in a face-centred tetragonal phase with high symmetry. It is interesting to note that the stabilization in any one of these two structures strongly depends on the size of the pnictogen atom and this structural competition plays a central role in determining which structural and magnetic phases are thermodynamically favourable. At low temperature, MnP and MnAs stabilize in orthorhombic and hexagonal structures, respectively. However, increasing of temperature MnAs is having structural phase transition from NiAs-to-MnP type structure at around 318-398\,K. Interestingly both MnSb and MnBi are stable in the hexagonal structure in the entire temperature range till the melting point. The atom positions of the MnN are Mn at $(0, 0, 0)$ and N at $(0,0,\,\frac{1}{2})$. In Hexagonal structure Mn atoms occupy the $2a$ Wyckoff site at $(0,\,0,\,0)$, while $X$ atoms reside on the $2c$ site at $\left(\frac{2}{3},\,\frac{1}{3},\,\frac{3}{4}\right)$. In the orthorhombic $Pnma$ (No.\ 62) structure, both Mn and $X$ atoms occupy the $4c$ Wyckoff positions. Mn is located at $(\tfrac{1}{4},\,y,\,z)$,
while $X$ resides at $(\tfrac{1}{4},\,y,\,z)$. Because of the lower symmetry of $B3_1$ compared to $B8_1$, the first and third coordinates of the above positions are free to relax and get distorted from the ideal high symmetric position in the hexagonal-to-orthorhombic structure.

\begin{table*}[t]
	\centering
	\caption{Calculated equilibrium structural parameters and magnetic moments of Mn$X$ ($X$ = N, P, As, Sb, Bi) compounds in their respective magnetic ground states. The optimized lattice constants ($a$, $b$, $c$), unit cell volumes (Volume), and magnetic moments ($M$) of the ground state structures are compared with corresponding available experimental values (shown in parentheses).}
	\label{tab:MnX}
	\setlength{\tabcolsep}{6pt}
	\begin{tabular}{lccccc}
		\hline\hline
		\textbf{Mn$X$} & $a$ (\AA) & $b$ (\AA) & $c$ (\AA) & Volume (\AA$^{3}$) & $M$ ($\mu_{\mathrm{B}}$/Mn) \\ 
		\hline\hline
		MnN (A-AFM)  & 4.19 (4.26\textsuperscript{a}) & 4.19 (4.26\textsuperscript{a}) & 4.10 (4.18\textsuperscript{a}) & 71.98 (75.86\textsuperscript{a}) & 2.90 (3.30\textsuperscript{a}) \\
		MnP (FM)  & 5.87 (5.90\textsuperscript{b}) & 5.21 (5.24\textsuperscript{b}) & 3.14 (3.24\textsuperscript{b}) & 96.29 (98.88\textsuperscript{b}) & 1.55 (1.61\textsuperscript{b}) \\
		MnAs (FM)  & 3.77 (3.72\textsuperscript{c}) & 3.77 (3.72\textsuperscript{c}) & 5.67 (5.70\textsuperscript{c}) & 69.57 (67.95\textsuperscript{c}) & 3.29 (3.43\textsuperscript{c}) \\
		MnSb (FM)  & 4.16 (4.14\textsuperscript{d}) & 4.16 (4.14\textsuperscript{d}) & 5.70 (5.77\textsuperscript{d}) & 85.42 (85.64\textsuperscript{d}) & 3.43 (3.50\textsuperscript{d}) \\
		MnBi (FM)  & 4.38 (4.27\textsuperscript{e}) & 4.38 (4.27\textsuperscript{e}) & 5.85 (6.09\textsuperscript{e}) & 97.19 (96.16\textsuperscript{e}) & 3.53 (3.85\textsuperscript{e}) \\
		\hline\hline
	\end{tabular}
	\\[2pt]
	\raggedright
	\textsuperscript{a}\cite{MnN} \quad
	\textsuperscript{b}\cite{MnP_MCE_2009} \quad
	\textsuperscript{c}\cite{MnAs} \quad
	\textsuperscript{d}\cite{MnSb} \quad
	\textsuperscript{e}\cite{MnBi}
\end{table*}

\section{Results and Discussion}
The MnN compound crystallize in distorted NaCl-type structure with tetragonal distortion of $c/a=0.98$. In its ground state, MnN adopts a collinear antiferromagnetic configuration with ferromagnetic layers having Mn with magnetic moment of $3.3\mu_B / Mn$ and Mn-N bond length around $\sim$2.0503\text{\AA} as shown in Fig.\ref{fig:struc}. The resulting base-centred tetragonal (BCT) cell corresponds to the true Bravais lattice in this tetragonal system, as the face-centred tetragonal (FCT) setting is not a standard cell choice within the known 14 Bravais lattices. It is experimentally observed\cite{ZHENG2023} that at high pressures AFM phase changes to FM phase around 35\,GPa where pressure induced FCT to cubic transition occurs simultaneously. The calculated exchange interaction parameters for MnN, obtained from the FP-LMTO method shows a clear competition between antiferromagnetic and ferromagnetic couplings. The first nearest neighbour interaction at $r_{ij}/a\approx0.7$ is strongly AFM (J<0), originating from exchange between Mn atoms in adjacent planes, whereas the third nearest neighbour interaction at $r_{ij}/a\approx1$ is strongly FM (J>0) and this is driven by Mn-N-Mn double exchange within the same plane. Beyond $r_{ij}/a\approx1.75$, the exchange parameters oscillate around zero with smaller values indicating negligible long-range magnetic coupling. The competition between the AFM and FM interactions not only stabilizes the A-AFM ordering but also accounts for its high N\'eel temperature, $T_N$ $(\sim 660K)$. From the calculated exchange energies mentioned above we have carried out finite temperature MC study and obtained the $T_N$ of 820\,K which is comparable with the experimental value of 660\,K\cite{MnN_struc_neel,MnN_neel}. The strong Mn–N–Mn exchange interactions preserve significant short-range magnetic correlations well above $T_{N}$, thereby limiting the increase in spin disorder across the transition. Consequently, the change in magnetic entropy remains low in the vicinity of $T_N$.
\begin{figure}[b!]
	\centering
	\includegraphics[scale=0.3]{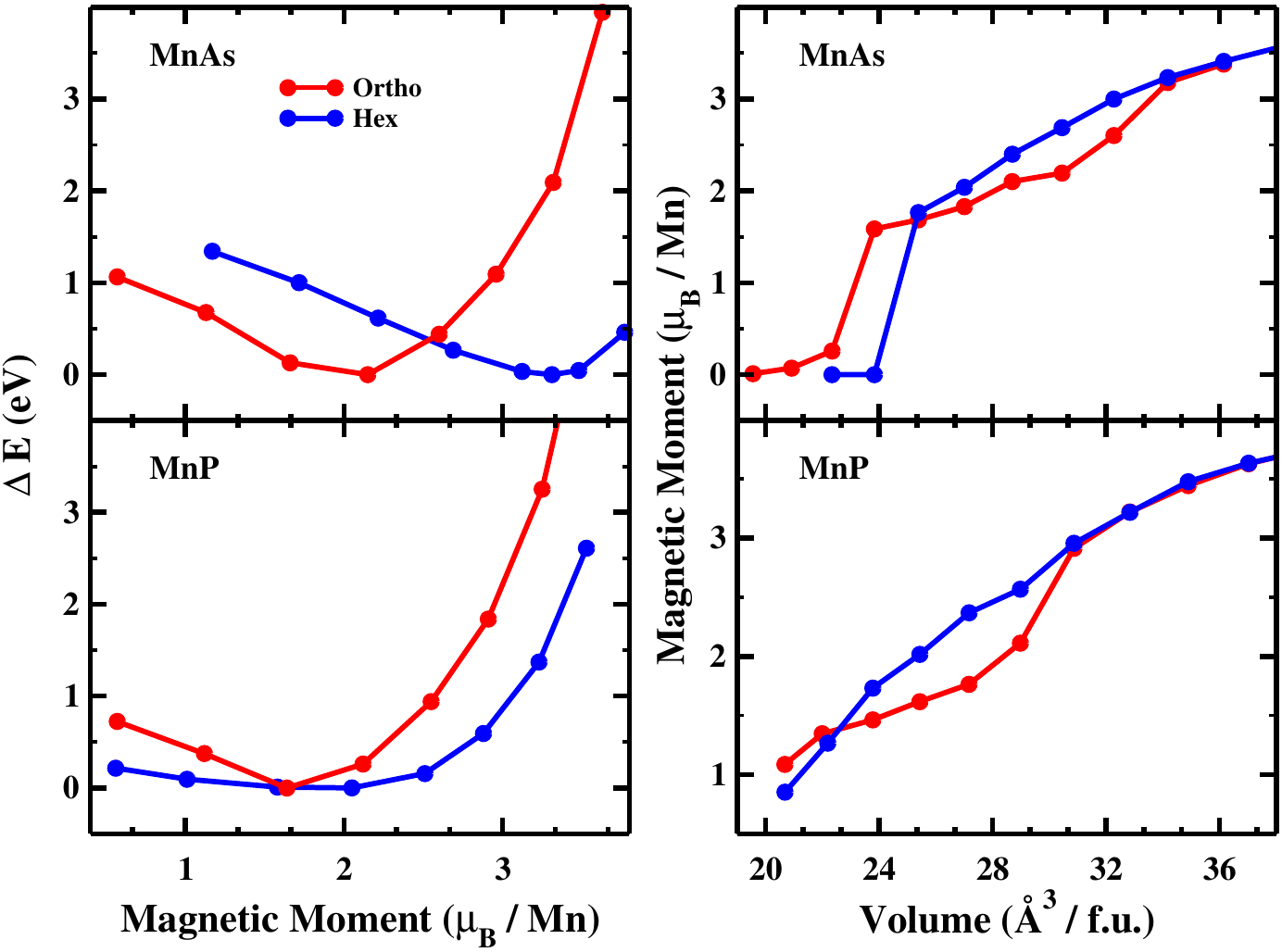}
	\caption{Fixed–spin–moment versus total energy curves with respect to lowest energy (left panel) and magnetic moment at Mn site changes with volume (right panel) for MnAs and MnP in the hexagonal (Hex) and orthorhombic (Ortho) structures. MnAs exhibits an energy minimum only in the higher moment regime, where Hex structure energetically  favoured, reflecting the large equilibrium volume that strengthens exchange splitting and suppresses the ortho distortion. In contrast, MnP shows its minimum at a low magnetic moment in its ground state Ortho structure and consistent with its reduced equilibrium volume where the $d$-band width enhanced by reduction in the bond length that stabilize low moment configuration.}
	\label{fig:fsmvvsm}
\end{figure}

\begin{figure*}[t!]
	\centering
	\includegraphics[scale=0.3]{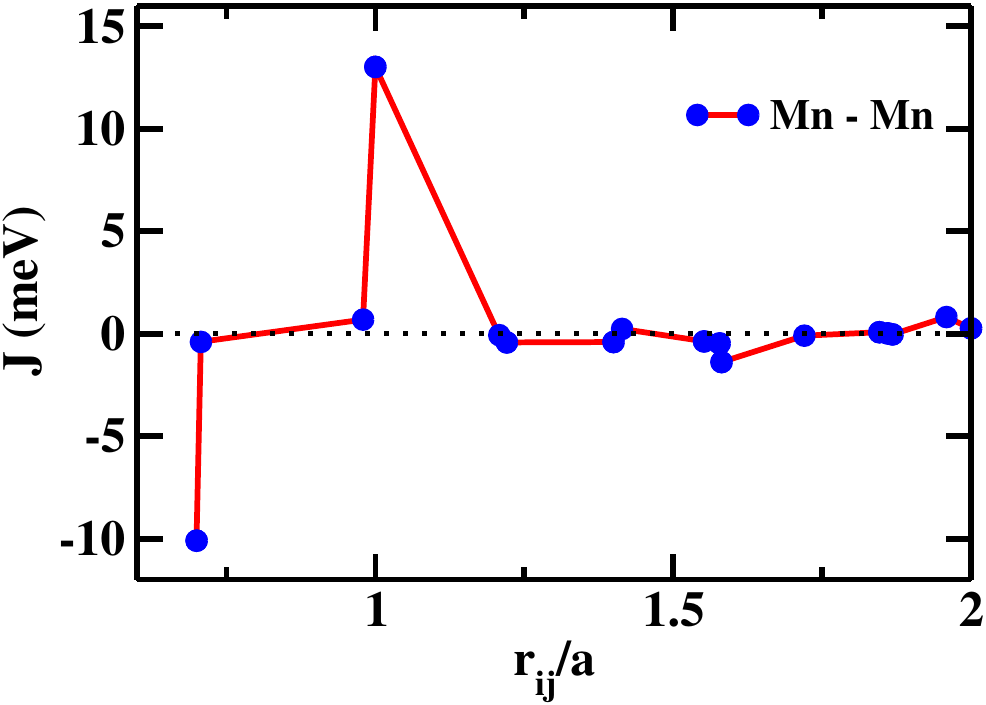}
	\includegraphics[scale=0.3]{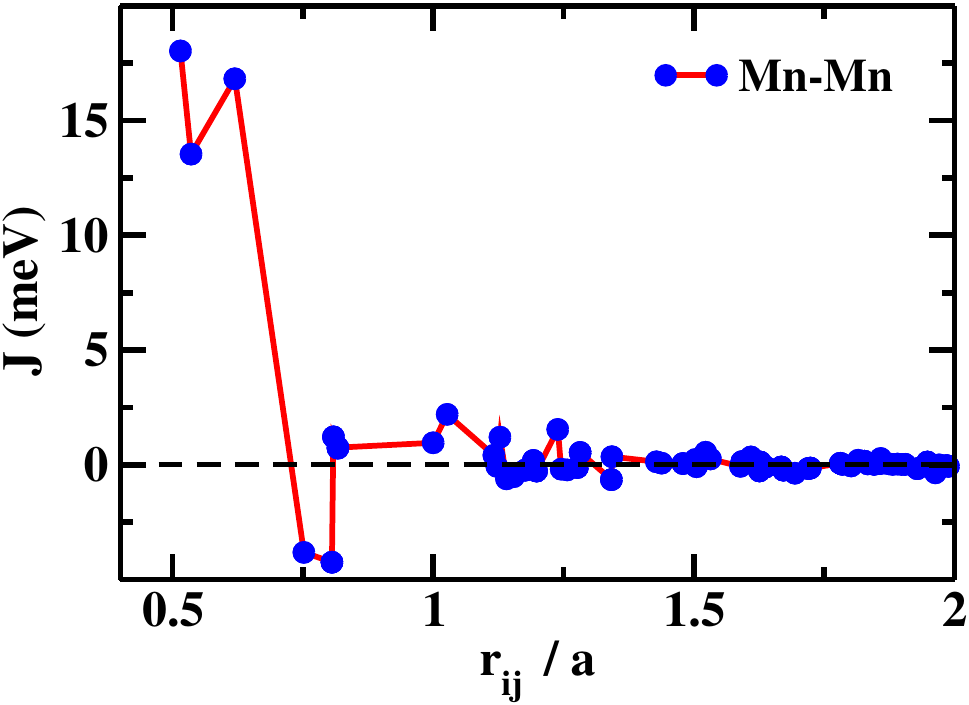}
	\includegraphics[scale=0.3]{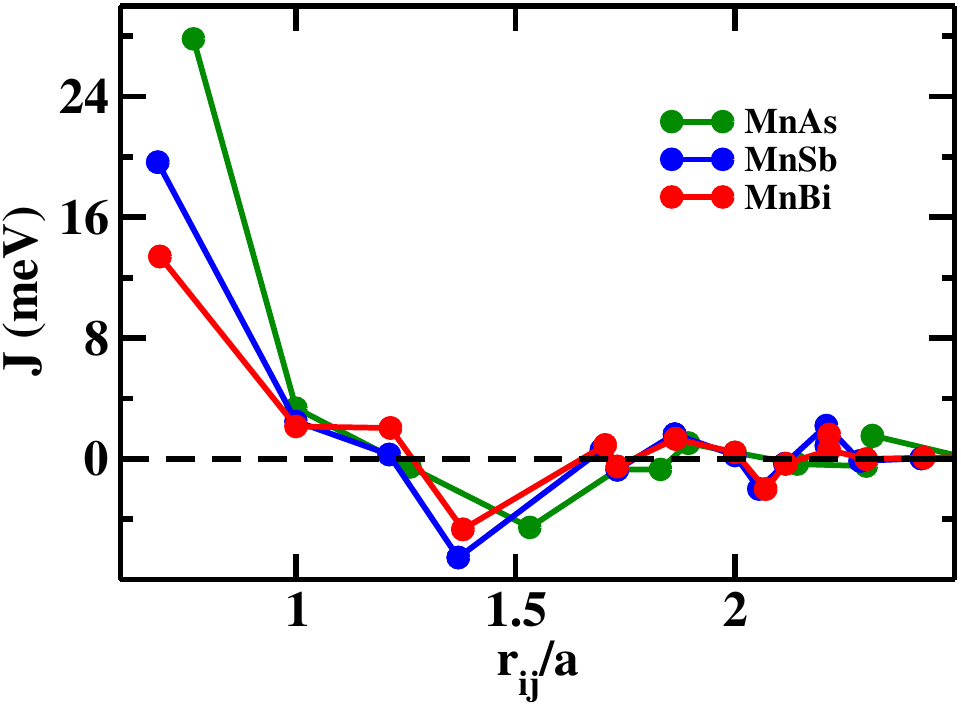}
	\caption{Exchange interaction between Mn atoms in MnN (left panel), MnP(middle panel) and Mn$Z$ ($Z$ = As, Sb, Bi) as a function of ratio between Mn-Mn distance($r_{ij}$) to the lattice parameter ($a$).}
	\label{fig:MnN-MnP}
\end{figure*}
The optimized lattice parameters along with corresponding experimental values are listed in the Table \ref{tab:MnX} for comparison. In MnP orthorhombic structure, Mn atoms are octahedrally coordinated by P atoms  and the Mn-P bond length ranges from 2.274\text{\AA}-to-2.488\text{\AA}. It is well established that the magnetic ground state of MnP significantly depend on temperature and magnetic field. MnP exhibits a sequence of magnetic phase transitions from paramagnetic-to-ferromagnetic state at 290\,K followed by a transition from ferromagnetic-to-helimagnetic phase around 50\,K.\cite{komatsubara1969,reisMnp008,MnP_MCE_2009,yamazakiNovelMagneticChiral2014} In the helimagnetic state, Mn spins rotate in the $ab$ plane with a propagation vector $Q_h=0.11$ along the $c$-axis\cite{reisMnp008,MnP_MCE_2009,xuMagnetic2017}. Furthermore, MnP is one of the few Mn-based compounds known to exhibit superconductivity (SC), emerging under high pressure conditions (7.5-8\,GPa). Recent work by Dissanayake \textit{et al.}\cite{HelicalMagneticState} has provided key insights into the pressure induced SC phase in MnP. The emergence of SC is closely linked with the critical magnetic transition from the helical-$c$ phase-to-the helical-$b$ phase, driven by pressure induced modifications in the exchange coupling landscape. In this study, first and second nearest neighbour exchange interactions are ferromagnetic in nature whereas, the third nearest neighbour exchange interaction becomes antiferromagnetic with increasing pressure, plays a decisive role in destabilizing the helical magnetic order and driving the emergence of superconductivity in MnP. Around 8\,GPa, where long-range magnetic order is nearly suppressed, SC appears with a transition temperature of $T_S$ = 20\,K. These findings suggest that the delicate competition between different magnetic exchange pathways under pressure especially involving long-range interactions are crucial for the unconventional SC pairing mechanism in MnP. It is experimentally known that the $c$-axis of MnP is the easy axis of magnetization which saturates at 7.5\,T magnetic field and the $b$-axis is intermediate axis, saturating around 40\,GPa, and the $a$-axis is the hard axis. We found that, the saturation magnetization as 104.66\,emu/g which is quite higher than that of Shapira \textit{et al.}\cite{shapiraPhaseTransitionsMnP1984} who obtained 84.3\,emu/g from vibrating-sample magnetometer and closer to experimentally reported value of Booth \textit{et al.}\cite{MnP_MCE_2009} 101.9\,emu/g from SQUID measurement.

The exchange interaction between Mn atoms are shown in Fig\ref{fig:MnN-MnP} as a function of ratio between Mn-Mn distance ($r_{ij}$) to the lattice parameter ($a$). The exchange interaction of the Mn-Mn pairs show long-range behaviour due to the metallic nature of this system. The first, second, and third nearest neighbours direct exchange interactions show large positive values (>10\,meV) and contribute to stabilizing the ferromagnetic phase. These three Mn-Mn pairs have a slightly different distance due to the octahedral distortion of the lattice. The variation in the Mn-P bond length within the octahedra alters the sign of the third-nearest-neighbour exchange coupling constant. On the other hand, the fourth, fifth, and sixth neighbours of the magnetic exchange coupling constants are attributed to the super-exchange type with fourth and fifth antiferromagnetic interactions having a value of around $-$4\,meV. It may be noted that the calculated exchange interactions from present study are in good agreement with previous theoretical works.\cite{HelicalMagneticState,tranTheoreticalPredictionLarge2021}

To probe the origin of high moment (HM-$\sim3\mu_B$) hexagonal and low moment (LM$\sim1.5\mu_B$) orthorhombic structures of MnAs and MnP, we have computed total energy vs fixed-spin-moment (FSM) and volume dependent magnetic moment M(V) curves for both these compounds in the hexagonal as well as orthorhombic structures. As shown in Fig.\ref*{fig:fsmvvsm}, the results reveal that the equilibrium volume plays an important role in deciding the magnetic moment of these systems irrespective of their crystal structures. For MnAs, the FSM curve shows a well-defined minimum at HM state in hexagonal phase, while the orthorhombic phase remains energetically unfavourable across all spin constraints. This reflects that the larger equilibrium volume of MnAs with reduced Mn-3$d$ bandwidth, enhances exchange splitting and stabilizes the HM configuration. In contrast, MnP shows its minimum at a LM moment in the orthorhombic phase, which arises from substantially smaller equilibrium volume with increased Mn-3$d$ bandwidth and reduced exchange splitting, thus favouring the LM state. so, the difference in magnetic moment at Mn site between MnP and MnAs does not reflect a spin-state transition but instead arises from itinerant meta magnetism. The continuous evolution of exchange splitting and minority-spin occupation, clearly visible in the orbital resolved DOS(Fig.4 in supplementary materials(SM)), confirms the itinerant nature of magnetism. The M(V) curves further demonstrate that both compounds follow the universal trend as LM at smaller volume and HM at larger volume (irrespective of the crystal structure) and this indicates crystal structure does not play any significant role. Implying that the contrasting magnetic ground states of MnAs and MnP arises primarily from their equilibrium volumes which are due to the size of the cation. 

Fig.\ref{fig:MnN-MnP} presents the calculated Mn$-$Mn exchange interactions $J$ in Mn$Z$ ($Z$ = As, Sb, Bi) as a function of normalized interatomic distances (ratio of Mn inter-atomic distances to the lattice constant, $a$). All these three compounds display a strong short-range ferromagnetic interaction, originating from direct Mn$-$Mn exchange at small separations. Among them, MnAs shows the largest nearest-neighbour $J$, consistent with its shorter Mn$-$Mn separations and the correspondingly stronger hybridized metallic bonding. As the Mn$-$Mn distance increases, the exchange interactions decay rapidly and oscillates around zero, indicating a crossover from direct-to-indirect exchange interaction. At sufficiently large separations, the $J$ values for MnSb and MnBi nearly coincide, reflecting the comparable contribution of their more diffuse pnictogen $p$ states in mediating indirect exchange. Across the series from MnAs-to-MnBi, the amplitude of first three direct exchange interactions $J$ systematically decrease due to the expansion of the lattice and the consequent reduction in Mn-Mn orbital overlap. Ferromagnetic MnAs exhibits the strongest short-range exchange interactions and the most localized magnetic moments in this series. The combined evolution of exchange strength, lattice expansion, and spin–orbit coupling governs the progression of magnetic ordering across Mn$Z$. The positive short-range exchange interactions stabilize the ferromagnetic ordering in Mn$Z$ systems.
\begin{table}[t]
	\centering
	\caption{Calculated orbital moment at Mn site for different magnetization axis and magneto-crystalline anisotropy energy for Mn$X$ systems.}
	\label{tab:mae}
	\begin{tabular}{lcccccc}
		\hline\hline
		& \multicolumn{2}{c}{Orbital moment [$\mu_B$/Mn]} & \multicolumn{2}{c}{$MAE$ (meV / f.u.)} \\
		\cline{2-3} \cline{4-5}
		Compound & {[001]} & {[100]} & Cal. & Exp./Theory*\\
		\hline
		MnN  &0.00 & 0.00 &0.09 & 0.13*\textsuperscript{a}\\ 
		MnP  &0.04 & 0.05 &0.22 & 0.08\textsuperscript{b}\\ 
		MnAs &0.02 & 0.03 &0.17 & 0.25\textsuperscript{c}\\ 
		MnSb &0.05 & 0.08 &0.18 & 0.21\textsuperscript{d}\\ 
		MnBi &0.13 & 0.17 &0.24 & 0.18\textsuperscript{e}\\ 
		\hline\hline
	\end{tabular}
	\\[2pt]
	\raggedbottom
	\textsuperscript{a}\cite{mnn_mae} \quad
	\textsuperscript{b}\cite{mnp_mae} \quad
	\textsuperscript{c}\cite{mnas_mae} \quad
	\textsuperscript{d}\cite{mae_mnsb} \quad
	\textsuperscript{e}\cite{mnbi_mae} \quad
\end{table}

\subsection{Magnetic Anisotropy Energy}
The magnetic anisotropy energy (MAE) for the Mn$X$ systems were calculated using the FP-LMTO method with spin-orbit coupling and orbital polarization (OP) correction. In this work, the MAE is defined as the difference between two self-consistently calculated fully relativistic total energies obtained for two different magnetization directions, $E_{001}-E_{100}$. To ensure good convergence, a large number of \textbf{k}-points were used in the IBZ, corresponding to 18,000\,-\,27,000 \textbf{k}-points in the full Brillouin zone. The results are summarized in Table\ref{tab:mae}. The orbital moment as well as the MAE increase when one impose OP correction to the scf calculations. It may be noted that OP correction properly account for Hund's rule and hence the calculated orbital moment will enhance with respect to SOC calculations and this results orbital moment are in good agreement with experiment\cite{fept_sir}. For MnN, the calculated MAE is 90\,$\mu$eV/f.u. with the easy axis lying in the basal plane. The experimental studies were carried out on MnP nanorods shows lesser MAE than our calculated result. Among the remaining pnictides MnBi shows largest MAE of 240\,$\mu$eV/f.u. while MnAs and MnSb shows comparable values of 170\,$\mu$eV/f.u. and 180\,$\mu$eV/f.u., respectively. The systematic trend of reduction of MAE is MnN-to-MnBi is mainly associated with reduction in SOC strength in Mn$X$ series. Okita \textit{et al.}\cite{mae_mnsb} observed the MAE of 210\,$\mu$eV/f.u. for MnSb at 77\,K and it is in good agreement with our predicted value. De Blois\textit{et al.}\cite{mnas_mae} experientially determined the MAE value for bulk MnAs at room temperature as well as 77\,K and the measured values are 390\,$\mu$eV/f.u. and 250\,$\mu$eV/f.u., respectively. Our calculated MAE for MnAs will be valid at ultra low temperature and hence, when we considered this fact our calculated MAE for MnAs is in good agreement with experimental measurement mentioned above. From the magnetization measurement for single crystal MnBi by Stutius \textit{et al.}\cite{mnbi_mae} determined the MAE of 180\,$\mu$eV/f.u. at 4.2\,K and other studies also show that MAE increase as temperature increases. The calculated orbital moments for different magnetization directions show element dependent variations across the Mn$X$ series. MnN exhibits a fully quenched orbital moment and other compounds display finite values depend on their Mn-$X$ hybridization and SOC strength. The shorter bond length of MnP leads to stronger Mn-P hybridization resulting reduced quenching of Mn 3$d$ states. The reduced hybridization due to longer bond length of MnAs, leaving the Mn orbital moment more quenched. The small discrepancy between our calculated MAE value and experimental results can be due to temperature effects, sample quality or sample stoichiometry.

\subsection{Magnetocaloric Effect}

\begin{figure}[b]
	\centering
	\includegraphics[scale=0.3]{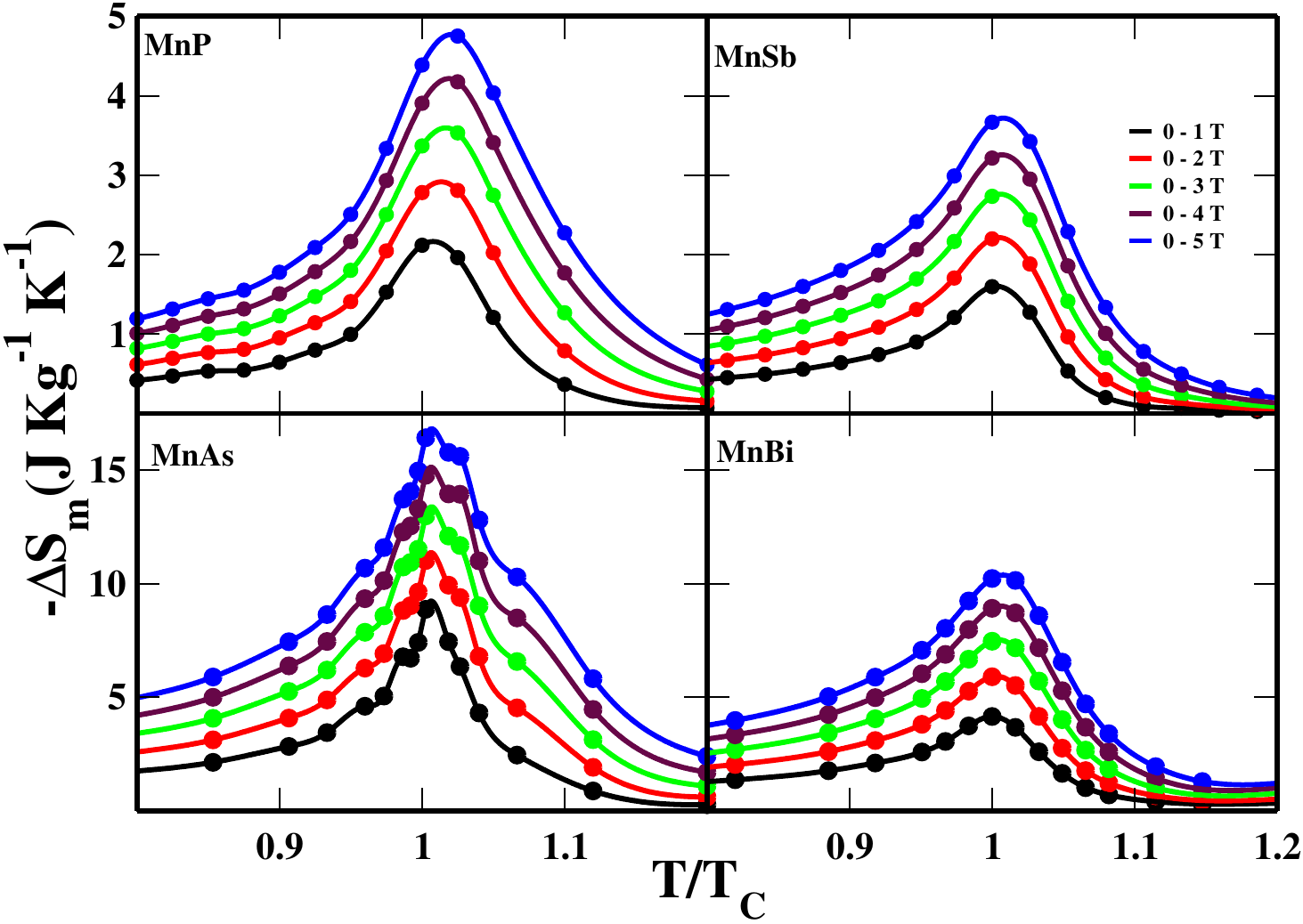}
	\caption{Magnetic entropy change as a function of normalized temperature for various applied external magnetic field for Mn$Y$ ($Y$=P, As, Sb, Bi)}
	\label{fig:all}
\end{figure}

Using the exchange interactions obtained from FP-LMTO method, we have calculated the magnetization change as a function of temperature using MC method. The magnetic entropy change as a function of normalized temperature (ratio of temperature to the magnetic transition temperature) for MnP, MnAs, MnSb, and MnBi under different applied magnetic fields are presented in Fig.\ref{fig:all}. All these compounds display well-defined peak near the Curie temperature, which is a characteristic signature of the magnetocaloric effect driven by the magnetic phase transition. The measurements were performed with the Mn$X$ crystals aligned along their easy magnetic $c$-axis which amplifies the magnetocaloric response due to enhanced magnetization changes near the magnetic transition temperature.
\begin{figure}[t]
	\centering
	\includegraphics[scale=0.4]{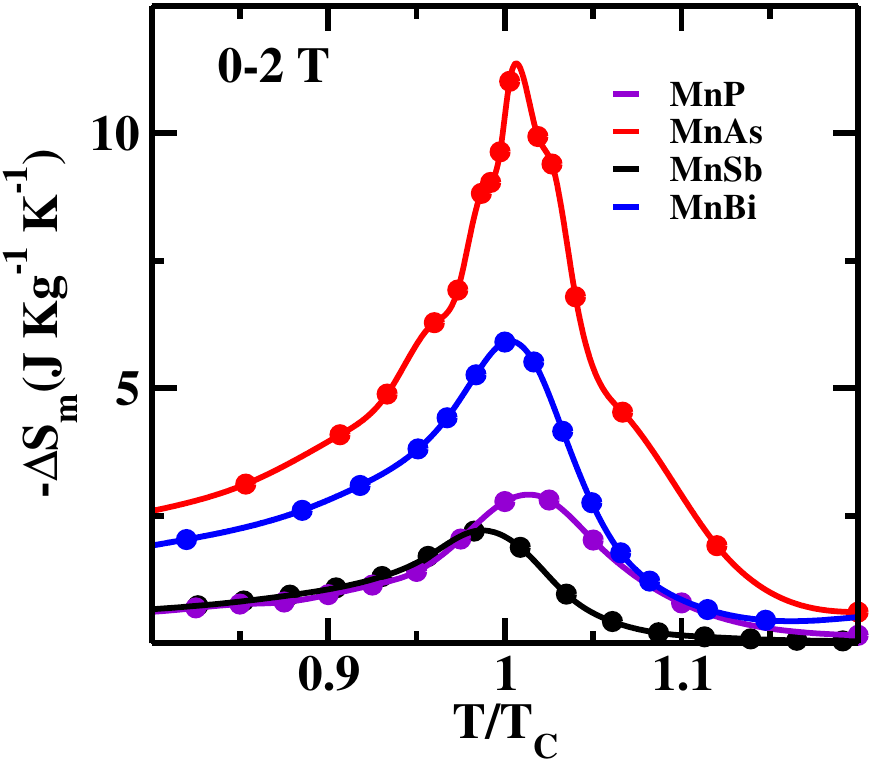}
	\caption{Comparison of magnetic entropy change as a function of normalized temperature at 2\,T of applied external magnetic field for Mn$Y$ ($Y$=P, As, Sb, Bi).}
	\label{fig:2T}
\end{figure}
The calculated magnetic entropy change for MnN in varying the applied magnetic field from 0-5\,T show a very low entropy change in the vicinity of magnetic phase transition as shown in SM. This low entropy change arises because of the transition occurring from a paramagnetic state to antiferromagnetic state. The $T_\mathrm{C}$ predicted for MnP by MFA is 680\,K, while the MC method yields 400\,K, which is slightly higher than the experimentally observed value of $290$\,K. Another theoretical study reported $T_\mathrm{C} = 665$\,K from MC and 948\,K by MFA~\cite{tranTheoreticalPredictionLarge2021}. The systematic overestimation of $T_\mathrm{C}$ in MFA with experiment and MC simulations mainly arises from non-accounting of thermal spin fluctuations and short-range correlations, whereas MC simulations explicitly account for these effects by statistically sampling spin configurations, making them more reliable for estimating magnetic phase transitions though they are computationally expensive.

As the field strength increases beyond 5\,T, the rate of change of magnetization with temperature diminishes sharply, leading to the saturation of the MCE. This saturation behaviour implies that, higher magnetic fields contribute progressively less to the magnetic entropy change ($\Delta S_\mathrm{M}$) compared to lower magnetic field. For a field change of 0-5\,T, the simulated $\Delta S_\mathrm{M}$ reaches a value of  $-$4.5\,J\,kg$^{-1}$\,K$^{-1}$ near $T_\mathrm{C}$, consistent with experimental observations.\cite{MnP_MCE_2009} Among the studied systems, MnAs exhibits the largest entropy change, reaching values above $-$16.43\,J\,kg$^{-1}$\,K$^{-1}$ at 0-5\,T, which reflects the strong first-order nature of its magnetic transition. Compared with experimental data, the calculated entropy change values are slightly lower, likely due to the underestimation of the discontinuity at the transition temperature within the DFT-based approach. In contrast, MnSb shows moderate entropy change with applied field and having  a peak value of $-$3.7\,J\,kg$^{-1}$\,K$^{-1}$, which agrees well with experiment. Interestingly, MnBi shows a smaller entropy change compared to MnAs, but the broad nature of its $\Delta S_\mathrm{M}(T)$ peak indicates a more gradual transition, yielding a maximum entropy change of $-$10.7\,J\,kg$^{-1}$\,K$^{-1}$. 

\begin{table}[b]
	\centering
	\caption{Calculated magnetic entropy change ($-\Delta S_{\text{M}}$) at a magnetic field variation of 0 - 5\,T and the magnetic transition temperature ($T_\text{C/N}$) from mean-field approximation (MFA), Monte-Carlo (MC), and experiment for Mn$X$ (X = N, P, As, Sb, Bi) are compared with available experimental values (shown in parentheses).}
	\label{tab:MnX_comparison}
	\setlength{\tabcolsep}{4pt}
	\begin{tabular}{lcccccl}
		\hline\hline 
		\rule{0pt}{3ex}	
		Compound & $-\Delta S_{\text{M}}$ & $T_\text{C/N}^{\text{MFA}}$ & $T_\text{C/N}^{\text{MC}}$ & $T_\text{C/N}^{\text{Exp}}$ & \\[3pt]		\vspace{4pt}
		& $(J\,kg^{-1}\,K^{-1})$ & (K) & (K) & (K) &  & \\ \hline 
		MnN  & $0.03 $ & 1063 & 820 & $\sim$660 & \\
		MnP  & $4.5 (4.1\textsuperscript{a})$  & 680 & 400 & 290 & \\
		MnAs & $16.4 (28\textsuperscript{b})$  & 535 & 370 & 318 & \\
		MnSb & $3.7 (3.66\textsuperscript{c})$ & 676.8 & 565 & 583 & \\
		MnBi & $10.7 $ & 693 & 610 & 628 &  \\ \hline\hline
	\end{tabular}
	\\[2pt]
	\raggedbottom
	\textsuperscript{a}\cite{MnP_MCE_2009}\quad
	\textsuperscript{b}\cite{mnas_mce}\quad
	\textsuperscript{c}\cite{MnSb}
\end{table}

For comparative analysis we have displayed the entropy change as a function of normalized temperature for all the ferromagnetic compounds considered in the present study at a magnetic field change of 0-2\,T as shown in Fig.~\ref{fig:2T}. In this figure MnP and MnSb having almost same value till $T/T_\mathrm{C} = 1$, while MnSb exhibits a slightly broader peak than MnP, which can be attributed to the stronger Mn$-$Mn exchange interaction. The strength of the magnetocaloric effect in Mn-based pnictides depends strongly on the anion species, with MnAs being the most promising candidate due to its large magnetic entropy change, whereas MnBi, despite a smaller peak  offers operational advantages through a broader working temperature window. The calculated magnetic entropy change ($-\Delta S_\mathrm{M}$) at an applied magnetic field of 0-5\,T and the transition temperature ($T_\mathrm{C}$) obtained from MFA, MC, and experiment are summarized in Table~\ref{tab:MnX_comparison}. The magnetic entropy change increases from MnN to MnAs, reaching a maximum value of approximately $-$16.4\,J\,kg$^{-1}$\,K$^{-1}$ for MnAs and this value is lower than the corresponding experimental value of $-$28\,J\,kg$^{-1}$\,K$^{-1}$. The enhancement of $-\Delta S_\mathrm{M}$ when one go from N-to-As can be attributed to the first-order nature of the magnetic transition in MnAs. In contrast, MnN exhibits a very small $-\Delta S_\mathrm{M}$ due to its predominantly antiferromagnetic character and weak spin-lattice coupling. The MC-derived $T_\mathrm{C}$ values show better agreement with experiment, validating the exchange interactions calculated in this study. Also, we have calculated the electronic entropy change $\Delta S_\mathrm{el}$ during the transition from antiferromagnetic-to-ferromagnetic for MnN and ferromagnetic-to-paramagnetic transition for Mn$Y$ ($Y$= P, As, Sb, Bi) systems using Sommerfeld approximation.\cite{sommerfeld} 
\begin{equation}
	\Delta S_{\mathrm{el}} \simeq \frac{\pi^{2}}{3} k_{B}^{2}\, T\, N(E_{F}), 
\end{equation}
where the $k_B$ is the Boltzmann constant and $N(E_{F})$ is the density of states at the Fermi level. The calculated $\Delta S_\mathrm{M}$ values are $-$12.18, $-$1.31, $-$4.41, $-$23.80, and $-$26.83 $Jkg^{-1}$\,K$^{-1}$ for MnN and Mn$Y$, respectively at corresponding magnetic transition temperatures. The increasing size of the pnictogens weakens Mn-$X$ hybridization and narrows the Mn-3$d$ bands, leading to a higher density of states at the Fermi level. This results increased electronic entropy change across the series. Overall, the variation in $\Delta S_\mathrm{M}$ and $T_\mathrm{C}$ with increasing cation size and structural properties demonstrates the delicate interplay between electronic structure and exchange interactions across the Mn$X$ series.

\subsection{Optical and Magneto-Optical Properties}
\begin{figure}[h]
	\centering
	\includegraphics[scale=0.27]{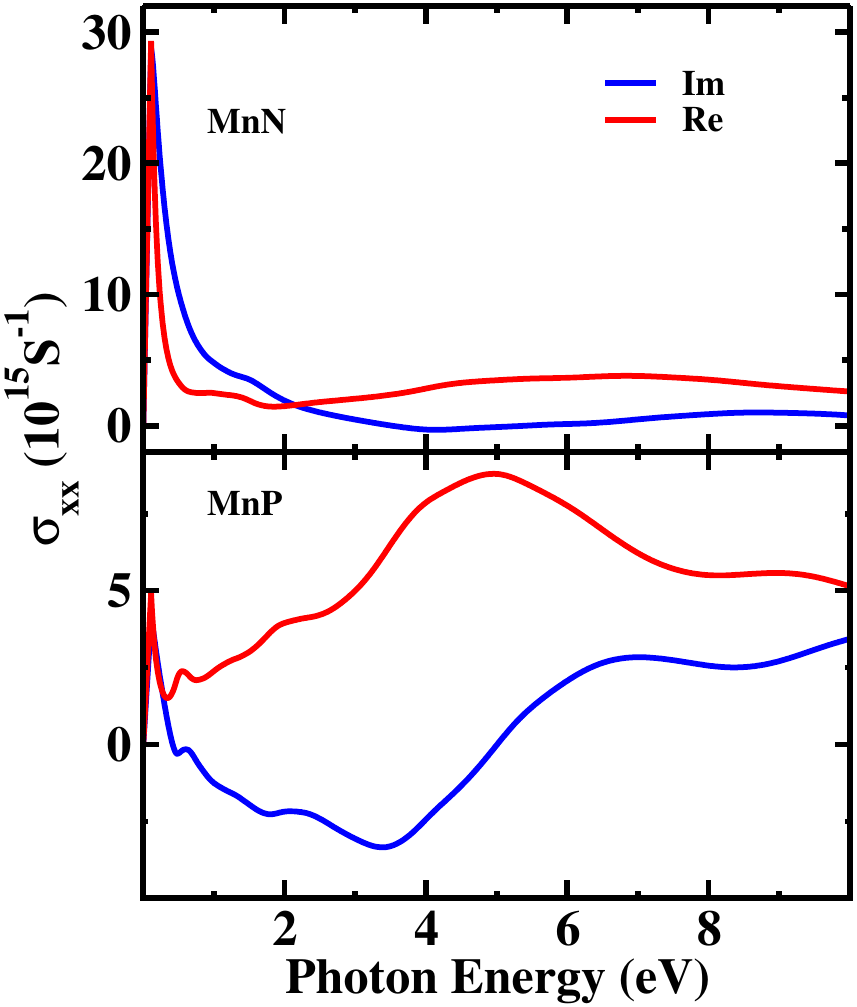}
	\includegraphics[scale=0.27]{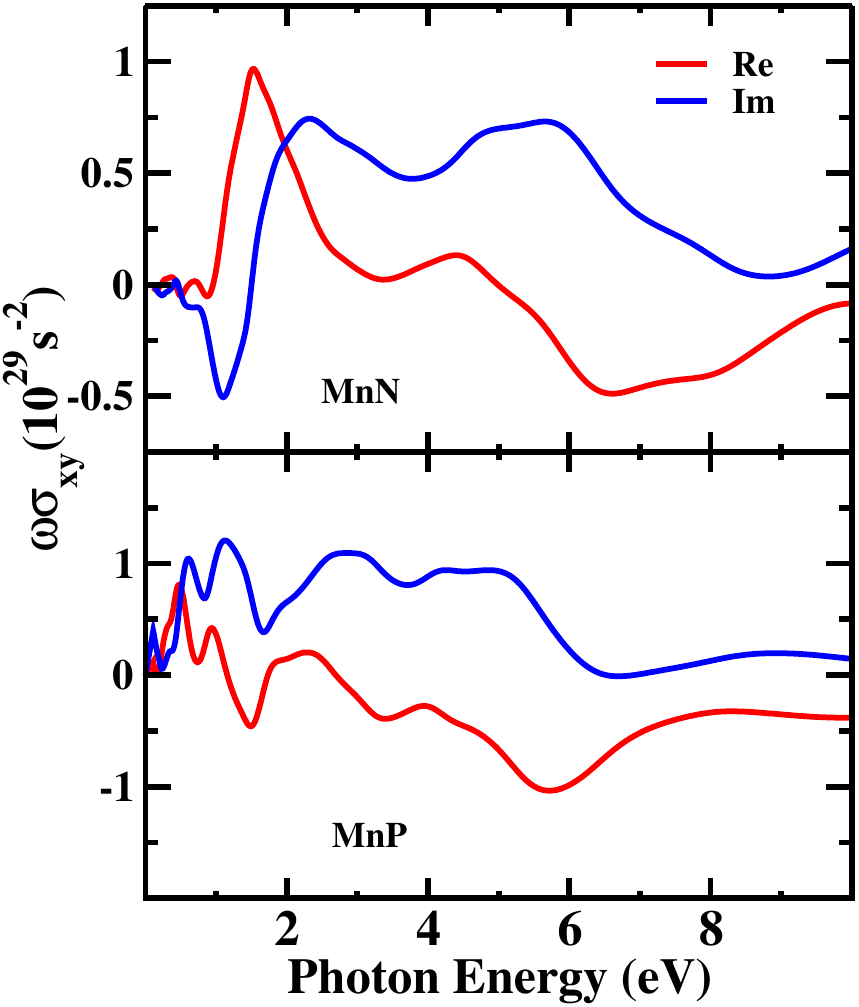}
	\caption{ Calculated complex diagonal (left panel) and off-diagonal (right panel) optical conductivity spectra  as a function of photon energy for MnN and MnP from FP-LMTO method.}
	\label{fig:cond2}
\end{figure}

\begin{figure}[h]
	\centering
	\includegraphics[scale=0.27]{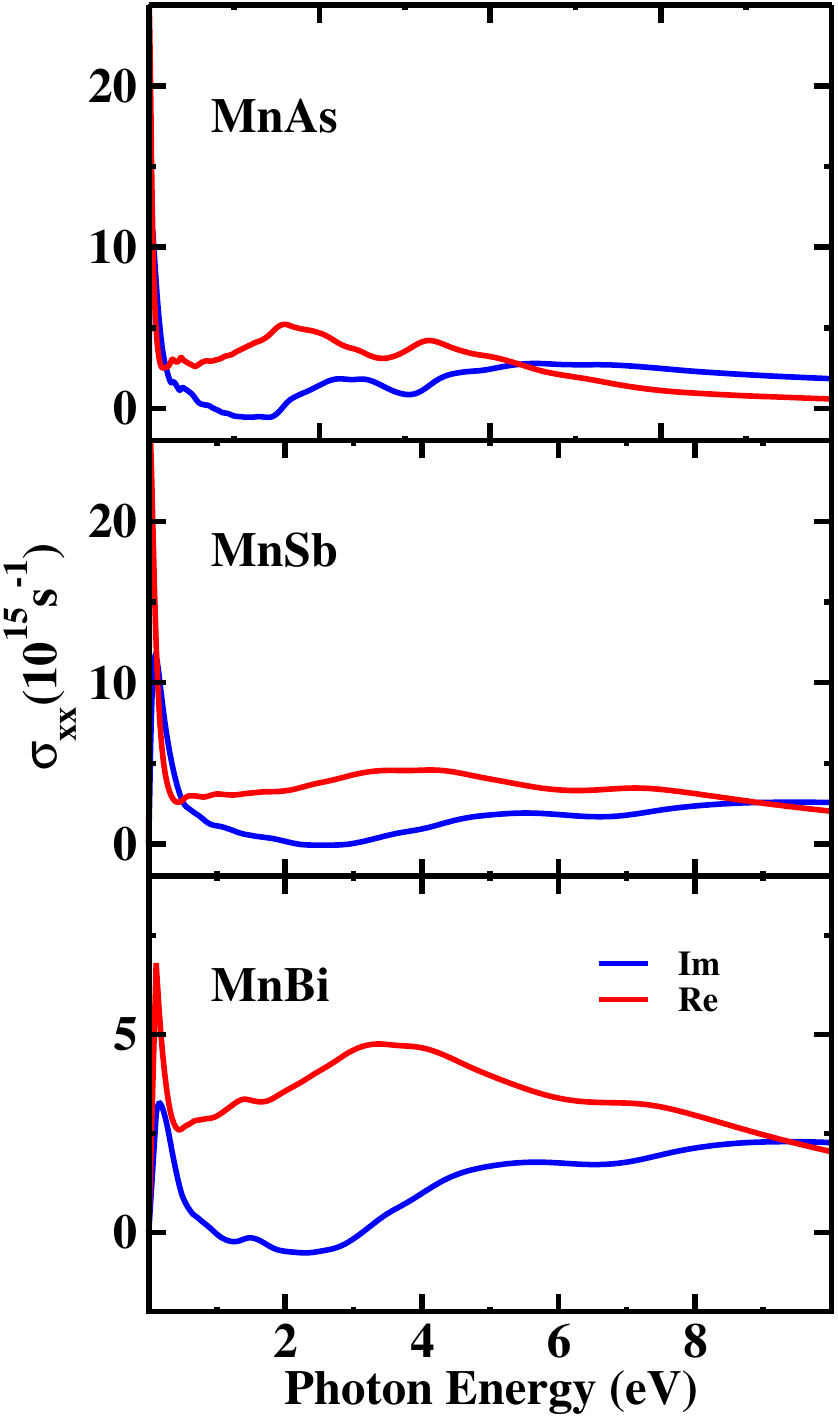}
	\includegraphics[scale=0.27]{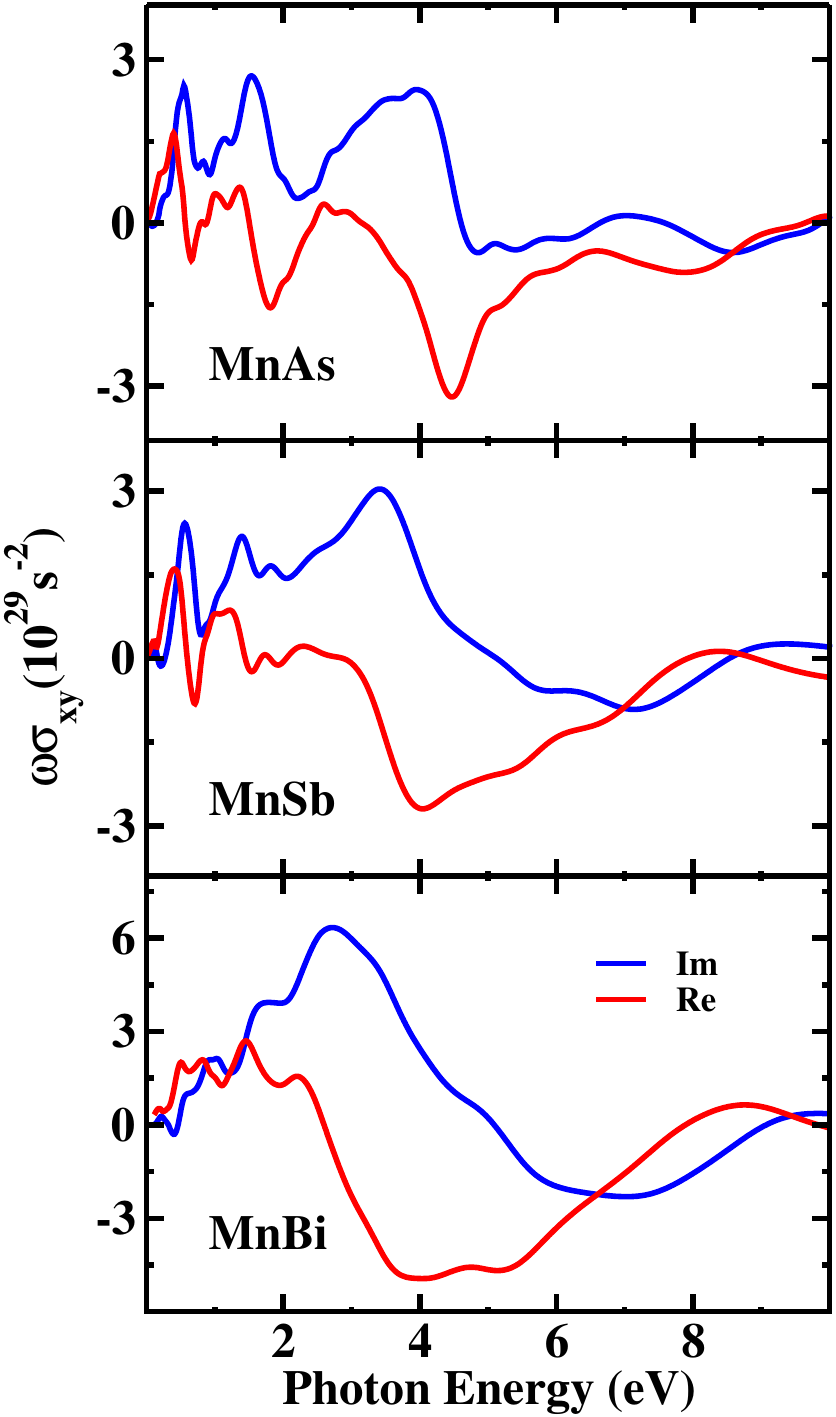}
	\caption{ Calculated complex diagonal (left panel) and off-diagonal optical conductivity spectra (right panel) as a function of photon energy for Mn$Z$ ($Z$ = As, Sb, Bi ) from FP-LMTO method.}
	\label{fig:cond3}
\end{figure}

\begin{figure}[b]
		\centering
		\includegraphics[scale=0.28]{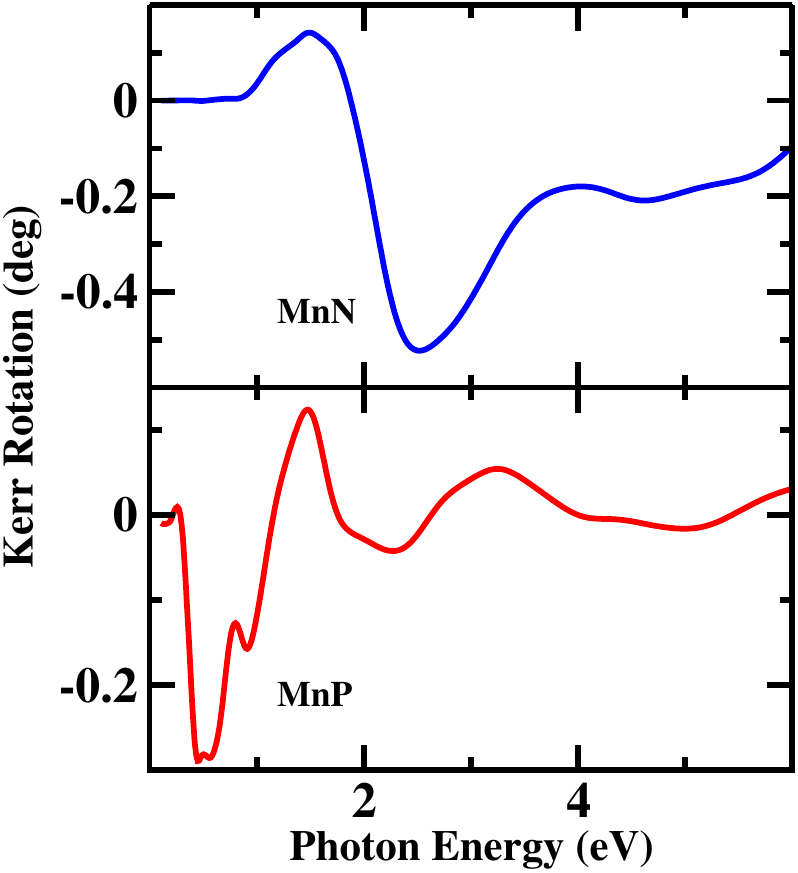}
		\includegraphics[scale=0.28]{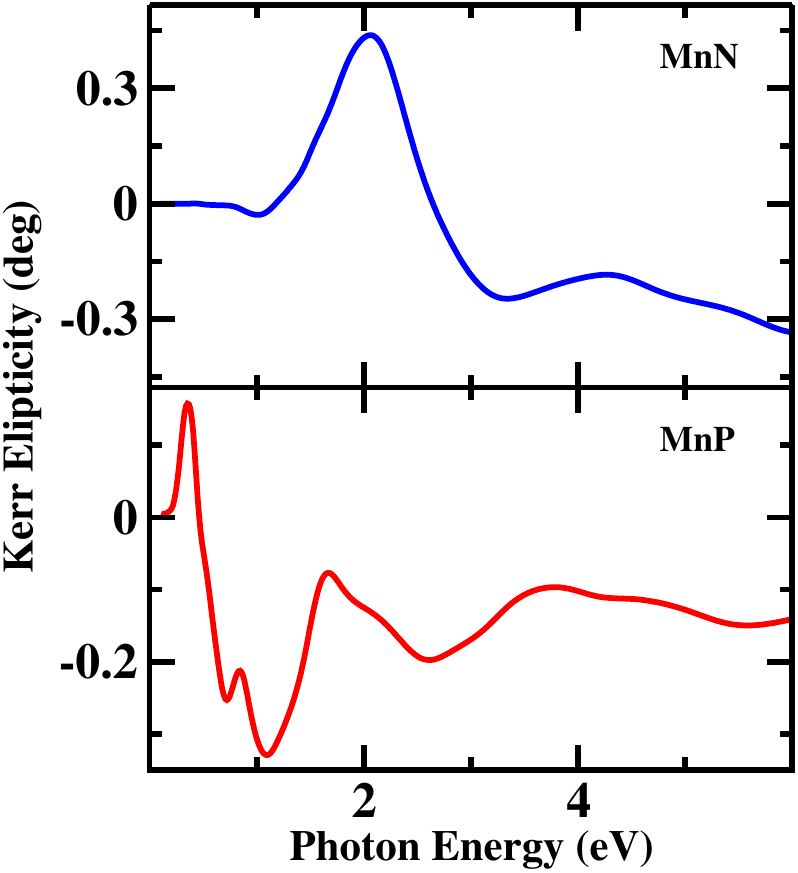}
		\caption{Calculated Kerr rotation (left panel) and Kerr ellipticity (right panel) as a function of photon energy for MnN, and MnP)}
		\label{kerr}
\end{figure}

The magneto–optical response of Mn$X$ ($X$ = N, P, As, Sb, Bi) compounds reveal rich sequence of spectral features that reflect their underlying electronic structure and the progressively increasing role of SOC across the pnictogen series. As shown in Fig.\ref{fig:cond2}, the diagonal optical conductivity of MnN and MnP are dominated by low-energy features, whereas MnAs, MnSb and MnBi in Fig.\ref{fig:cond3} display broader peaks extending well into the higher energy regime. This evolution signals enhanced Mn-$X$ hybridization and stronger relativistic interactions as the pnictogen atomic number increases. The off-diagonal part of optical conductivity, which governs the MO activity also follows a systematic chemical trend, MnN and MnP exhibit relatively small amplitudes, while MnSb and especially MnBi show much larger values. For MnN, the Kerr ellipticity Fig.\ref{kerr} attains comparatively large magnitudes in the low-energy regime, indicating significant absorptive magneto–optical activity. By contrast, the Kerr rotation remains almost negligible throughout the entire photon-energy window examined. This decoupling between rotation and ellipticity implies that the real and imaginary parts of the off-diagonal optical conductivity nearly cancel, suppressing the phase difference required to generate a substantial Kerr rotation. This behaviour highlights that the magneto-optical response in Mn$X$ compounds are determined not solely by the strength of SOC but also by the interplay between Mn 3$d$ - X $p$ hybridization. 

\begin{table}[t]
	\centering
	\caption{Calculated unscreened plasma frequencies by using WIEN2k code($\omega_p$) of Mn$X$ ($X$ = N, P, As, Sb, Bi) compounds along the principal crystallographic directions. The values of $\omega_{xx}$, $\omega_{yy}$, and $\omega_{zz}$ correspond to the diagonal components of the plasma tensor, and the average plasma frequency (Avg $\omega_p$) is obtained by taking the mean of these three components.}
		\setlength{\tabcolsep}{4pt}
\label{tab:plasma_freq}
\begin{tabular}{ccccc}
	\hline\hline
	Compound & $\omega_{xx}$ (eV) & $\omega_{yy}$ (eV) & $\omega_{zz}$ (eV) & Avg $\omega_P$ (eV) \\
	\hline
	MnN & 7.34 & 7.34 & 7.16 & 7.28 \\
	MnP & 2.35 & 3.47 & 2.52 & 2.90 \\
	MnAs & 4.53 & 4.53 & 4.86 & 4.64 \\
	MnSb & 4.60 & 4.60 & 4.96 & 4.72 \\
	MnBi & 3.47 & 3.47 & 4.41 & 3.81 \\		
	\hline\hline
\end{tabular}
\end{table}

Progressing to heavier pnictides, MnAs and MnSb show intermediate responses. Among the systems considered in the present study, MnAs displays a finite Kerr rotation across the investigated photon-energy range, although its overall magnitude is smaller than that of MnBi as shown in Fig.\ref*{kerr3}. The Kerr ellipticity in MnAs presents a more structured profile with distinct sign reversals and reaching a maximum of $-$0.55$^\circ$ at 1.8\,eV, reflecting the moderate SOC strength of As. In MnSb, both Kerr rotation and ellipticity attain higher amplitudes, with the rotation reaching about $-$0.6$^\circ$ near 1.3\,eV and the ellipticity approaching $-$0.8$^\circ$ at 3.8\,eV. Faraday spectra also shows a clear progressive trend from N-to-Bi by reaching a maximum Faraday rotation of $7.5\times10^5 deg/cm$ at 1.8\,eV and Faraday ellipticity of $14\times10^5 deg/cm$ at 3.5\,eV for MnBi. These features point to stronger relativistic contributions and enhanced Mn\,-\,Sb band mixing relative to MnAs

\begin{figure}[h]
	\centering
	\includegraphics[scale=0.28]{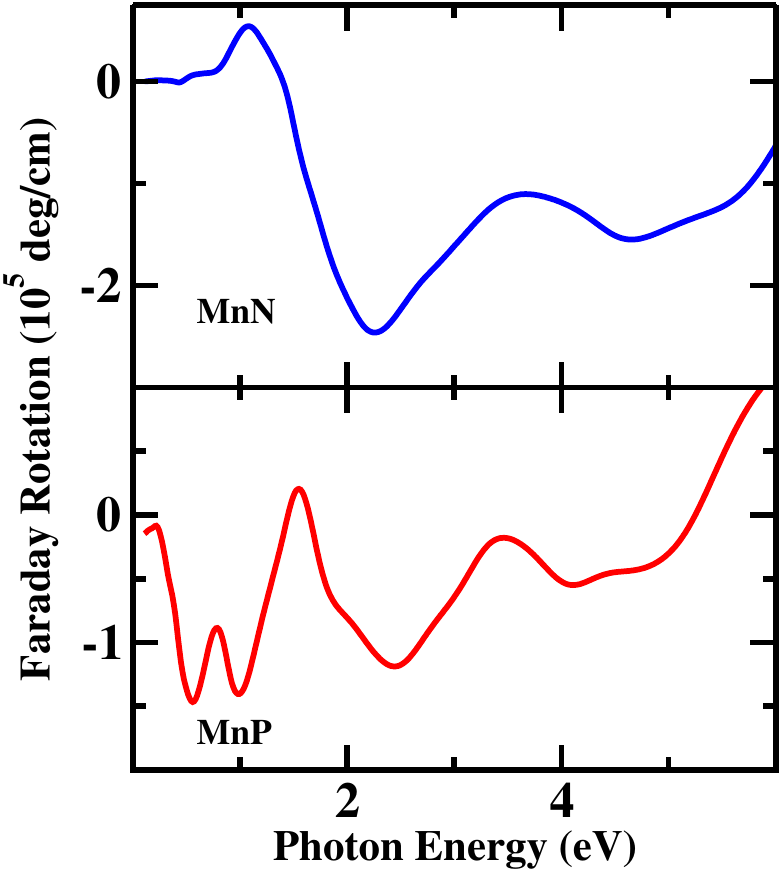}
	\includegraphics[scale=0.28]{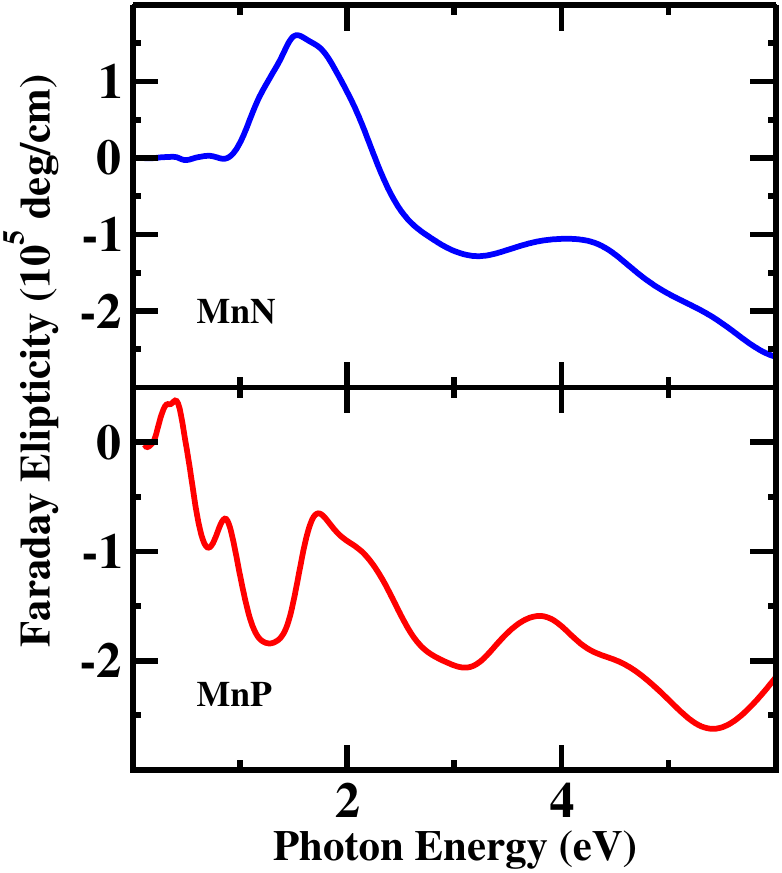}
	\caption{Calculated Faraday rotation (left panel) and Faraday ellipticity (right panel) as a function of photon energy for MnN and MnP}
	\label{fara}
\end{figure}
Among the Mn$X$ series, MnBi clearly emerges as the most magneto-optically active compound. The Kerr rotation exhibits the largest amplitude, approaching $-$1.2$^\circ$ at 1.6\,eV, while the Kerr ellipticity displays broad and strongly enhanced negative peaks at higher energies. The pronounced response of MnBi is a direct consequence of the strong SOC introduced by Bi, along with large magnetic moment at Mn site maximize exchange splitting and corresponding band splitting and amplifies the off-diagonal optical conductivity.

The microscopic origin of pronounced MO behaviour in MnBi can be traced to the cooperative effect of three ingredients, the large spin moment at Mn site, the strong SOC of Bi, and the pronounced hybridization between Mn 3$d$ and Bi 6$p$ states. In MnBi, these ingredients act synergistically, the Mn exchange field and Bi SOC affect the same hybridized electronic states, thereby producing a maximal off-diagonal response. Analysis of densities of states confirm that Mn $d$ and Bi $p$ derived features closely track one another, and that the spin-polarised states near the Fermi level enable large transverse $p-d$ transitions driven by SOC. In general, the Kerr rotation and ellipticity stem from the off-diagonal elements of the optical conductivity tensor, which are determined by interband dipole matrix elements and the energy spacing of Kohn-Sham states. In MnBi, the dominant Kerr rotation peak at 1.8\,eV coincides with a maximum in the absorptive part of off-diagonal conductivity. These correspondences reveal how the hybridized Mn-Bi states leave their imprint on the optical conductivity and ultimately on the Kerr spectra. The discrepancy between the different theories are due to neglecting orbital polarization correction and not accounting intraband contribution to the optical dielectric tensor. In general, inclusion of intraband contribution to the optical dielectric tensor significantly enhances the diagonal component up to 2\,eV. This increased diagonal response suppresses the Kerr rotation in the low-energy region. The calculated MO spectra of Ravindran \textit{et.al}\cite{ravindranMnX} reported earlier having a noticeable difference from present study and this is associated with anti-site configuration used earlier ($X$ atoms in the basal plane, are stacked along $c$-axis and Mn atoms in the octahedral voids).   
\begin{figure*}
	\centering
	\includegraphics[width=0.45\textwidth]{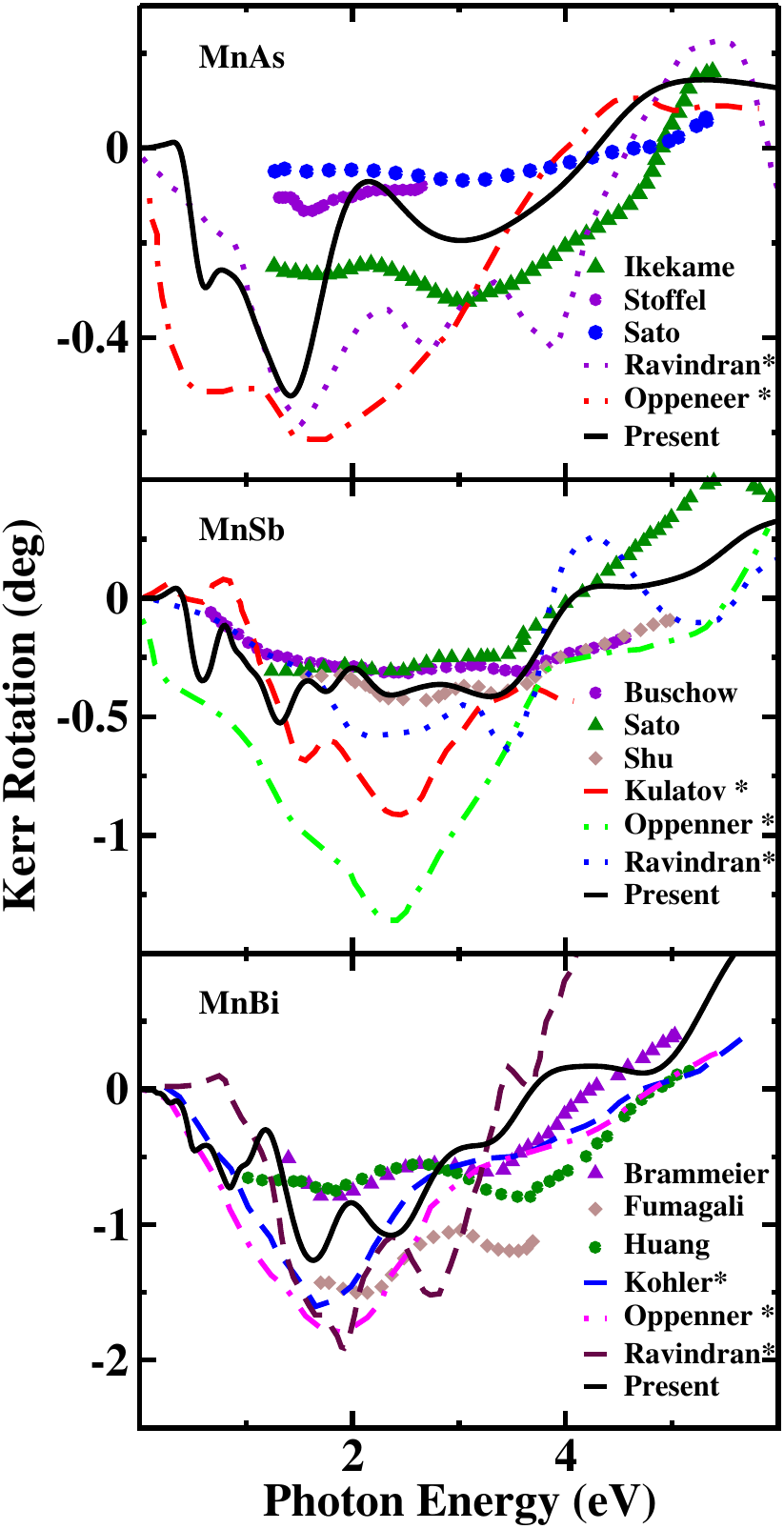}
	\includegraphics[width=0.45\textwidth]{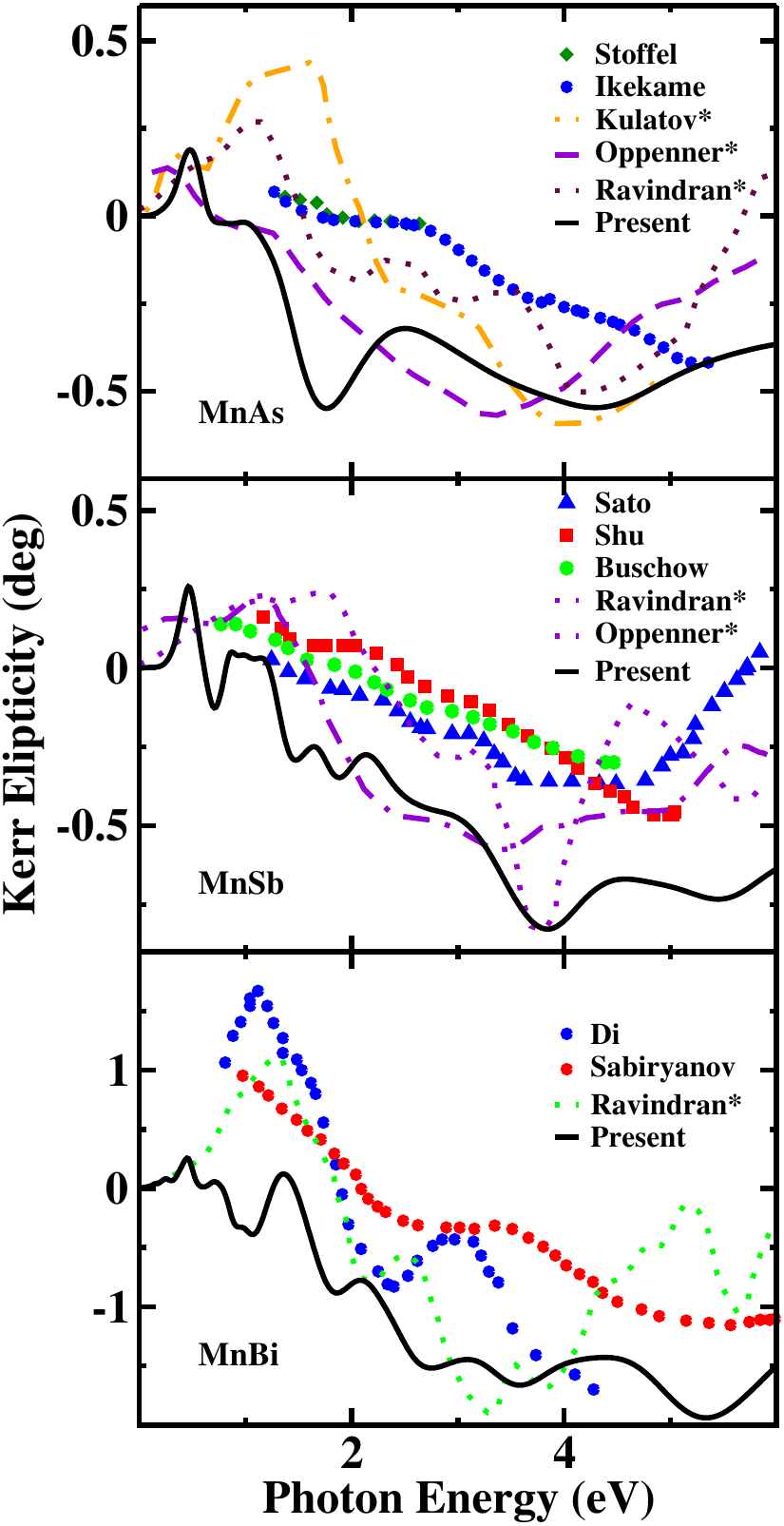}
	\caption{calculated Kerr rotation and Kerr ellipticity as a function of photon energy for Mn$Z$ ($Z$ = As, Sb, Bi). For MnBi, the experimental spectra are taken from Brammier \textit{et al.}\cite{brammeier2004magneto}, Fumagali \textit{et al.}\cite{fumagalli1996}, Huang \textit{et al.}\cite{huang} Sabiryanov \textit{et al.}\cite{sabiryanov} and Di \textit{et al.}\cite{di1996optical} and theoretical spectra from Oppenner \textit{et al.}\cite{oppeneer2001magneto}, Ravindran\textit{et al.}\cite{ravindranMnX} and  Kohler \textit{et al.}\cite{kohler1997calculated}. For MnSb, the experimental spectra are taken from Sato \textit{et al.}\cite{sato_bulk_mnsb}, Shu \textit{et al.}\cite{shu} and Bushow \textit{et al.}\cite{buschow} and the theoretical spectra from Kulatov\textit{et al.}\cite{Kulatov1995}. For MnAs the experimental spectra are taken from Stoffel \textit{et al.}\cite{stoffel_mnas} and Ikekame\textit{et al.}\cite{ikekame} * represents theory }
	\label{kerr3}
\end{figure*}

The experimental MO measurements of MnBi thin-films show two distinct Kerr-rotation peaks, a dominant low-energy peak near 1.8\,eV and a weaker higher energy peak near $\sim$3.2\,eV. The low-energy peak originates from interband transitions of Mn 3$d$ electrons hybridized with Bi 6$p$ states. The strong SOC of the Bi further splits and shifts unoccupied Bi-$p$ states toward the Fermi level, enhancing the probability of such low-energy transitions. The origin of the higher energy Kerr feature around 3.2\,eV has been more controversial. Early band structure calculations by K\"ohler and K\"ubler\cite{kohler1997calculated} did not reproduce this feature for stoichiometric MnBi. They proposed that this higher energy peak arises from interstitial oxygen ($MnBiO_{0.5}$), enabling O-$p$ down - Mn $d$ down spin transitions near $\sim$3.4\,eV. Supporting this view, Harder \textit{et al.}\cite{harder} observed only the low-energy peak in MnBi grown under ultra-high vacuum condition. In contrast, Brammeier \textit{et al.}\cite{brammeier2004magneto} reported a weak but distinct feature at $\approx$3.2\,eV in nearly oxygen free single crystals. 
\begin{figure*}
	\includegraphics[width=0.45\textwidth]{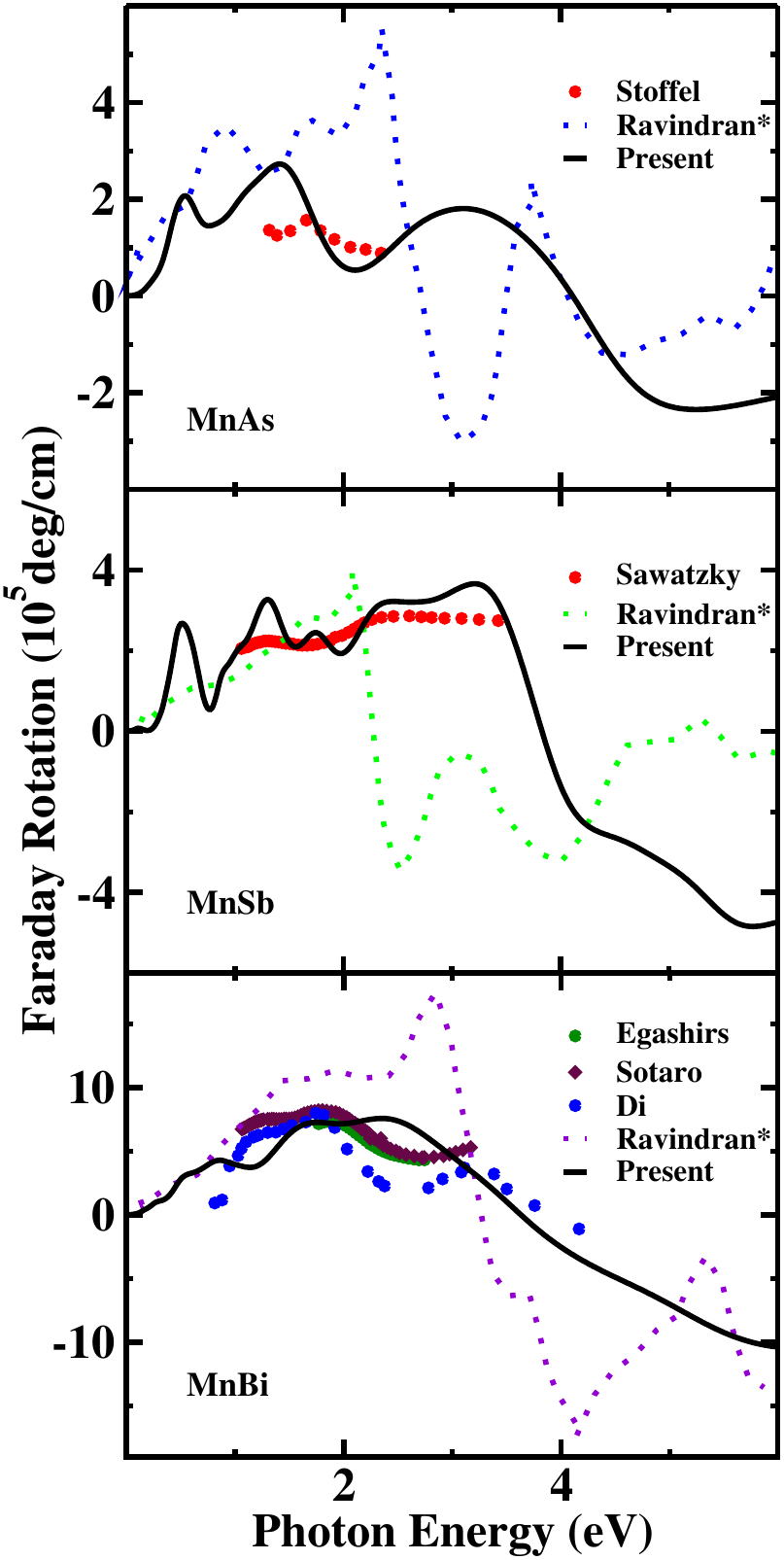}
	\includegraphics[width=0.45\textwidth]{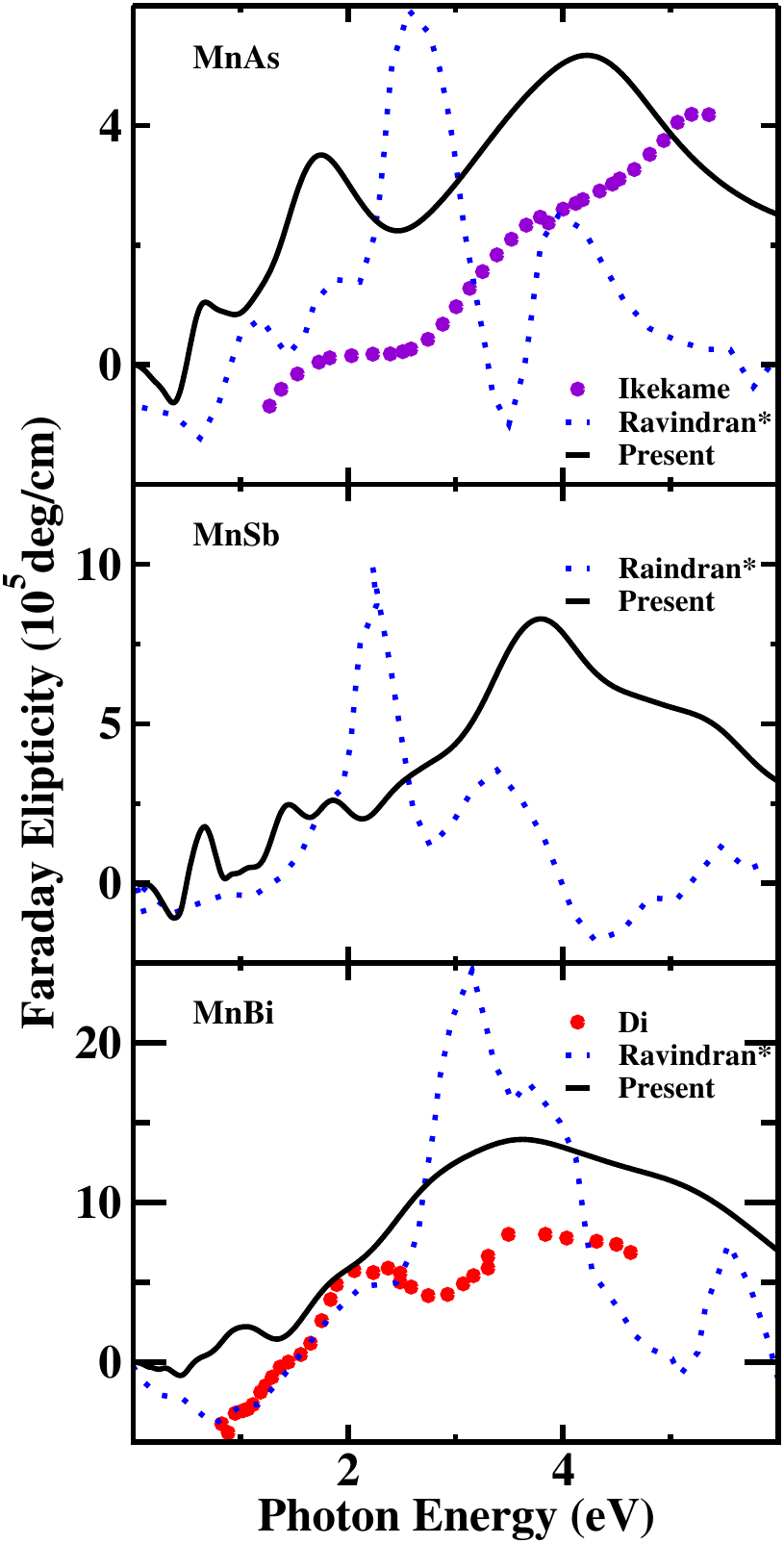}
	\caption{Calculated Faraday rotation (left panel) and Faraday ellipticity (right panel) as a function of photon energy for Mn$Z$ ($Z$ = As, Sb, Bi) For MnBi, the experimental spectra are taken from Di \textit{et al.}\cite{di1996optical}, Sotaro \textit{et al.}\cite{esho1974optical} and Egashira \textit{et al.}\cite{egashira}. For MnAs, Stoffel \textit{et al.}\cite{stoffel_mnas} and Ikekame\textit{et al.}\cite{ikekame}. * represents theory}
	\label{fara3}
\end{figure*}

They further argued that a thin oxide overlayer would mainly modify the peak amplitude rather than generate a new spectral feature. It may be noted that the photoelectron spectroscopy shows oxygen valence states $\approx$6\,eV below Fermi level too deep to produce a Kerr response at $\approx$3.2\,eV. Taken together, these findings indicate that the intensity of high-energy peak depend on surface conditions or stoichiometry, the existence of a Kerr feature near 3.2\,eV is most likely an intrinsic property of high-quality MnBi. Our results demonstrate that the -1.2$^\circ$ Kerr rotation at 1.5\,eV is an intrinsic Mn-Bi electronic structure and the higher energy Kerr feature is also intrinsic but appears in our calculations at 2.6\,eV. Both features originate from zero-crossings in the imaginary part of the diagonal optical conductivity near 1.6\,eV and 2.6\,eV, combined with pronounced peaks in the imaginary part of the off-diagonal conductivity at the same energies. Thereby reconciling earlier discrepancies between experiment and theory.

\begin{table}[b]
	\centering
	\caption{Comparison of spin ($m_{\text{spin}}$) and orbital ($m_{\text{orb}}$) magnetic moments for Mn$X$ compounds using XMCD sum rule and from spin-polarized calculation.}
	\label{xmcd}
	\setlength{\tabcolsep}{3pt}
	\begin{tabular}{llcc}
		\hline
		\hline
		\rule{0pt}{3ex}	
		\textbf{Mn$X$} & \textbf{Method} & \textbf{$m_{\text{spin}}$ ($\mu_B$/Mn)} & \textbf{$m_{\text{orb}}$ ($\mu_B$/Mn)} \\
		\hline
		\hline
		\rule{0pt}{2ex}	
		\multirow{2}{*}{MnN}   & Sum rule  & 2.7046 & -0.0042 	\\
	                       	   & DFT & 2.8994 & -0.0069 \\ \hline
	    		\rule{0pt}{2ex}	
		\multirow{2}{*}{MnP}   & Sum rule  & 1.4637 & 0.0181 \\
     						   & DFT & 1.5520 & 0.0166 \\ \hline
     		    \rule{0pt}{2ex}	
   		\multirow{3}{*}{MnAs}  & Sum rule  & 3.0947 & 0.0252 \\
							   & DFT & 3.1729 & 0.0188  \\
						       & Exp\textsuperscript{a}& 3.4300 & 0.0240 \\  \hline
				\rule{0pt}{2ex}	
		\multirow{2}{*}{MnSb}  & Sum rule  & 3.2297 & 0.0329  \\
		 					   & DFT & 3.4207 & 0.0358 \\ \hline
		 		\rule{0pt}{2ex}	
		\multirow{3}{*}{MnBi}  & Sum rule  & 3.3951 & 0.1063 \\
							   & DFT & 3.8629 & 0.1003 \\
							   & Theory\textsuperscript{b} & 3.5800 & 0.0841 \\ \hline
	\end{tabular}
		\\[2pt]
		\raggedbottom
	    \textsuperscript{a}\cite{mnas_xmcd} \quad
		\textsuperscript{b}\cite{mnbiorbitalmoment} \quad
\end{table}
 
\begin{figure}
	\centering
	\includegraphics[angle=270,scale=0.25]{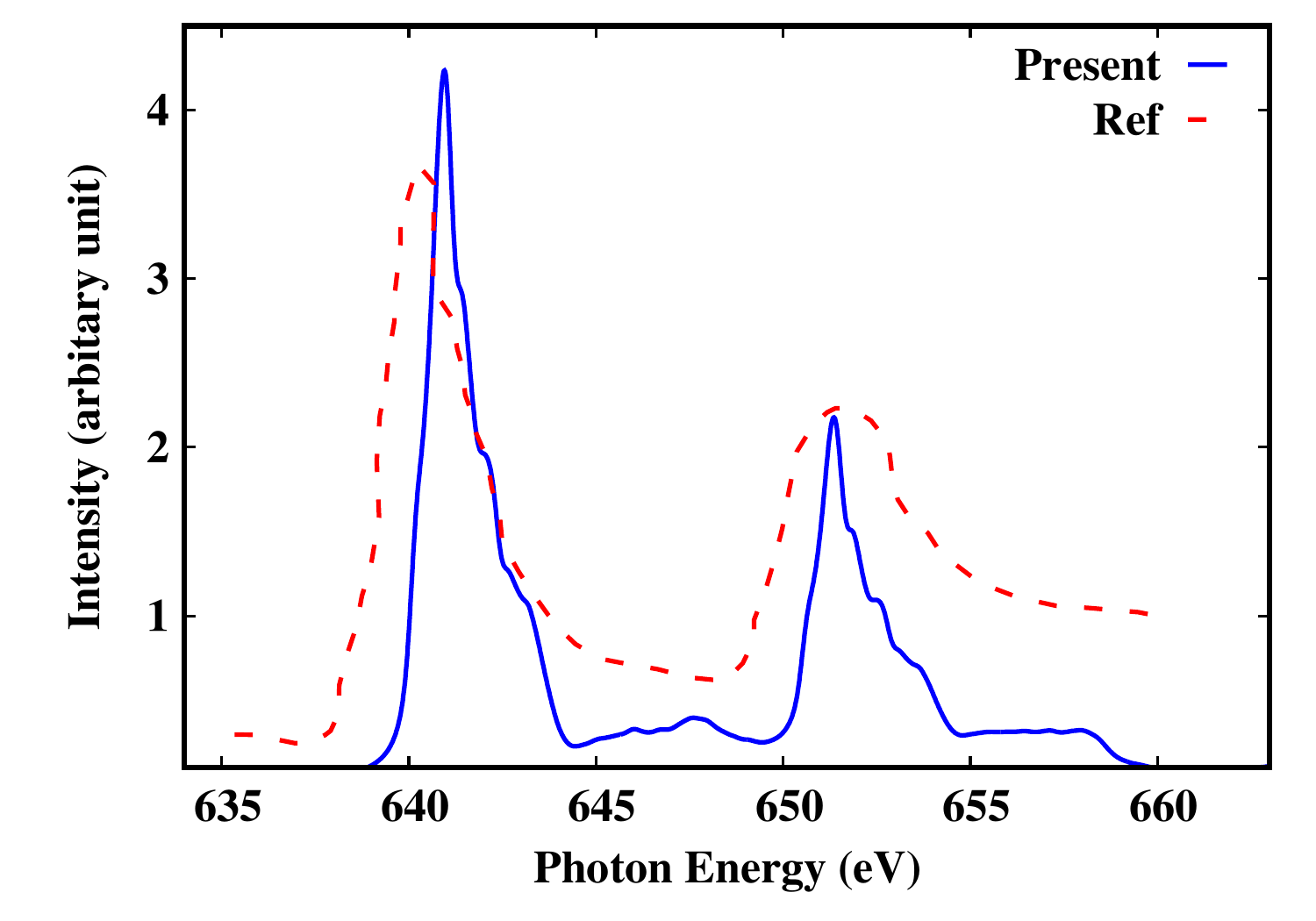}
	\caption{Calculated Mn $L_{2,3}$-edge XAS and XMCD spectra for MnAs. Experimental spectrum taken from Okabayashi \textit{et al.}\cite{xas_mnas_ref}}
	\label{fig:mnasxmcd}
\end{figure}

The X-ray absorption spectroscopy at the Mn $L_{2,3}$-edge were analysed using APW+lo method. The Mn $L$-edge spectra, arising from 2$p\rightarrow3d$ dipole allowed transitions, exhibit the strong spin polarization characteristics of 3$d$ states. Comparison of calculated spectra of Mn site at different MnX compounds show that variations in Mn–$X$ hybridization produce subtle modifications in the Mn $L$-edge profile. At the Mn $L_{2,3}$-edge, the intense lower energy feature correspond to the $L_3  (2p_{3/2} \rightarrow 3d$)  transitions, while the higher energy peak originates from the $L_2 (2p_{1/2} \rightarrow 3d$) transitions. XMCD spectra at the Mn $L_{2,3}$-edge for Mn$X$ were analysed using magneto-optical sum rules to quantitatively extract the spin and orbital magnetic moments posses at the Mn site. The comparison of experimental and calculated XAS spectra for MnAs compound are presented in Fig.\ref{fig:mnasxmcd}. For the other compounds, XAS and XMCD spectra are given in SM. 
The spin and orbital moments were evaluated from the XMCD sum rules\cite{sumrule}. These values are in good agreement with the magnetic moments obtained from self-consistent DFT calculations and are tabulated in Table\ref{xmcd}. This confirms the accuracy of the simulated spectra and underlying electronic structure. Across Mn$Y$ ($Y$ = P, As, Sb, Bi) series, the Mn magnetic moments exhibit a systematic evolution. The spin moment increases approximately from 1.46$\mu_B$ for MnP to 3.86$\mu_B$ for MnBi, reflecting progressively weaker Mn-$X$ $p-d$ hybridization and enhanced localization of Mn 3$d$ states for system with heavier pnictogens. The orbital moment, nearly quenched in MnN (-0.004$\mu_B/Mn$), becomes progressively significant in MnBi (0.10$\mu_B/Mn$), consistent with the unquenching of orbital angular momentum driven by strong SOC in the heavy Bi atom. These results validates the increasing SOC influence across the Mn$X$ series, which directly correlates with their enhanced magneto crystalline anisotropy and magneto-optical activity.

\section{Conclusion}
In summary, a comprehensive DFT and MC-based investigation of the Mn$X$ ($X$ = N, P, As, Sb, Bi) pnictides has been carried out systematically. The calculated results reveal a progressive evolution of magnetism and spin-orbit driven effects across Mn$X$ series, governed primarily by Mn-$X$ hybridization and the increasing strength of spin-orbit coupling when go from N-to-Bi. Structural stability and spin-state energetics show that MnP stabilizes in a low moment orthorhombic phase, whereas MnAs favours a high moment hexagonal structure, reflecting a strong magneto elastic coupling between lattice volume and magnetic order. The calculated exchange interactions indicate that MnN exhibits competing ferromagnetic(FM) and antiferromagnetic(AFM) couplings, resulting in an AFM ground state with a relatively high $T_N$. On the other hand, the heavier pnictides stabilize in a FM ground state with an increasing $T_C$ as the pnictogen atomic size increases. The entropy changes calculated from MC simulations are in good agreement with available experimental results. Among this series, MnAs is the most efficient magnetocaloric material with large magnetic entropy change due to its sharp first-order transition and magnetic transition around room temperature. Similarly, MnBi also show promising magnetic entropy change values with high $T_C$.  
\newline
Magneto-optical calculations demonstrate a progressive enhancement of Kerr and Faraday responses from MnN to MnBi. Among Mn$X$, MnBi exhibits the largest Kerr rotation as a direct consequence of strong Mn 3$d$ exchange splitting and large Bi spin-orbit coupling. Our results demonstrate that the higher energy Kerr feature is intrinsic and appears at slightly lower energy than experimentally observed value. The systematic increase in the MO effect when go from N-to-Bi is due to the enhancement of magnetic moment at Mn site and increased spin-orbit coupling of pnictogens. The excellent agreement of the calculated orbital moment from $L_{2,3}$-edge spectra with XMCD sum-rule analysis and those obtained from spin-orbit coupling included DFT calculations confirm the reliability of the calculated electronic structure and the accuracy of the simulated Mn $L_{2,3}$-edge spectra. Our results provides a unified picture linking the magnetocaloric and magneto-optical functionalities of Mn-based pnictides to their underlying magnetic and electronic structures. Overall, this study establishes the structural and magnetic evolution across the Mn$X$ series as the pnictogen size increases, highlighting their versatility and strong potential for applications in spintronic, optical, and solid-state refrigeration technologies.

\begin{acknowledgments}
Jayendran S is grateful for the financial support from the Research Council of Norway (NAMM project), Central University of Tamil Nadu and SCANMAT Centre for providing supercomputer facility.
\end{acknowledgments}

\bibliographystyle{apsrev4-2}
\bibliography{MnX}
\end{document}